\documentclass[a4,notitlepage,12ptm]{jedm}

\usepackage{nameref}

\usepackage{floatrow}
\usepackage[pdftex]{graphicx}
\graphicspath{{./}}
\DeclareGraphicsExtensions{.eps,.pdf,.jpeg,.png}
\usepackage{verbatim}

\usepackage[cmex10]{amsmath}
\usepackage{mathtools}
\usepackage{bm}
\makeatletter
\g@addto@macro\normalsize{%
  \setlength\abovedisplayskip{5pt}
  \setlength\belowdisplayskip{5pt}
  \setlength\abovedisplayshortskip{7pt}
  \setlength\belowdisplayshortskip{7pt}
}
\g@addto@macro\small{%
  \setlength\abovedisplayskip{5pt}
  \setlength\belowdisplayskip{5pt}
  \setlength\abovedisplayshortskip{7pt}
  \setlength\belowdisplayshortskip{7pt}
}
\makeatother
\usepackage{amssymb}

\usepackage{array}
\usepackage{booktabs}
\usepackage[font=small]{caption}
\usepackage[caption=false,font=footnotesize]{subfig}
\usepackage{floatrow}
\usepackage{etoolbox}

\usepackage{url}

\newcommand{\squeezeup}{\vspace{-0mm}}
\newcommand{\norm}[1]{\left\lVert#1\right\rVert}

\AtBeginEnvironment{align}{\par\small}
\AfterEndEnvironment{align}{\normalsize}
\AtBeginEnvironment{align*}{\par\small}
\AfterEndEnvironment{align*}{\normalsize}

\begin{document}
\title{Next-Term Student Performance Prediction: A Recommender Systems Approach}


\date{}

\author{%
    {\large Mack Sweeney}\\
        George Mason University\\
        msweene2@gmu.edu \\ 
        ~\\
    {\large Huzefa Rangwala}\\
        George Mason University\\
        rangwala@cs.gmu.edu\\  
    ~\\
    {\large Jaime Lester}\\
        George Mason University\\
        jlester2@gmu.edu \\ 
    ~\\
    {\large Aditya Johri}\\
        George Mason University\\
        ajohri3@gmu.edu 
 }

\maketitle

\AtBeginEnvironment{tabular}{\par\footnotesize}
\AfterEndEnvironment{tabular}{\normalsize}
\BeforeBeginEnvironment{table}{\par\footnotesize}
\AfterEndEnvironment{table}{\normalsize}

\begin{abstract}

\vspace{-5mm}
An enduring issue in higher education is student retention to successful
graduation. National statistics indicate that most higher education institutions
have four-year degree completion rates around 50\%, or just half of their
student populations. While there are prediction models which illuminate what
factors assist with college student success, interventions that support course
selections on a semester-to-semester basis have yet to be deeply understood. To
further this goal, we develop a system to predict students' grades in the
courses they will enroll in during the next enrollment term by learning patterns
from historical transcript data coupled with additional information about
students, courses and the instructors teaching them.

We explore a variety of classic and state-of-the-art techniques which have
proven effective for recommendation tasks in the e-commerce domain. In our
experiments, Factorization Machines (FM), Random Forests (RF), and the
Personalized Multi-Linear Regression model achieve the lowest prediction error.
Application of a novel feature selection technique is key to the predictive
success and interpretability of the FM. By comparing feature importance across
populations and across models, we uncover strong connections between instructor
characteristics and student performance. We also discover key differences
between transfer and non-transfer students. Ultimately we find that a hybrid
FM-RF method can be used to accurately predict grades for both new and returning
students taking both new and existing courses. Application of these techniques
holds promise for student degree planning, instructor interventions, and
personalized advising, all of which could improve retention and academic
performance.

\end{abstract}


\section{Introduction}

An enduring issue in higher education is student retention to successful
graduation \cite{NCES2013}. The 2001 National Research Council report
\cite{NRC2001} identified the \textit{critical need} to develop innovative
approaches to help higher-education institutions retain students, facilitate
their timely graduation, and ensure they are well-trained and workforce ready in
their field of study. The question then, is how can we increase student
retention and knowledge acquisition? A highly referenced report by Vincent Tinto
suggests that research on retention has made great advances, but effective
implementation efforts by universities are few and far between
\cite{tinto_research_2006}. Some notable exceptions were detailed by The
Educational Trust \cite{kevin_carey_choosing_2005}. There are three key
characteristics of successful retention-increasing programs. The first two are
student engagement and a genuine emphasis on teaching and learning. The third,
and the one we focus on most in this study, is \textit{effective utilization of
new data systems to monitor student progress and identify key aspects of student
success}.

The ability to predict student grades in future enrollment terms provides
valuable information to aid students, advisors, and educators in achieving the
mutually beneficial goal of increased student retention. This information can be
used to help students choose the most suitable majors, properly blend courses of
varying difficulty in a semester's schedule, and indicate to advisors and
educators when students need additional assistance. Early identification of
at-risk students is a key aspect of preventing them from becoming discouraged
and dropping out \cite{grayson1998identifying}. Incorporation of next-term grade
predictions in early-warning systems can increase the number of such students
correctly identified, potentially increasing retention rates significantly.
Furthermore, successful prediction models will yield valuable insights into what
factors impact student success across various subpopulations. These insights can
inform policy decisions to increase student engagement to further increase
retention.

In this paper, we develop a system to predict students' course grades for the
next enrollment term in a traditional university setting \footnote{Source
available at: \url{https://github.com/macks22/ntsgp}}. Students take courses
over a sequence of academic terms. During each term, students enroll in one or
more courses and earn letter grades in the range A-F for each. Our dataset
consists of grades from previous terms, which we call historical transcript
data, coupled with student demographics, course characteristics, and instructor
information. Given this data, our task is to predict the grades students will
obtain in the courses they will take in the next term.

With this problem formulation, the next-term student grade prediction problem
becomes quite similar to recommendation problems such as rating prediction and
next-basket recommendation. Throughout the years, a variety of methods have been
used for recommendation in the e-commerce domain \cite{bell_bellkor_2007}.
These recommender methods typically fall into one of two categories, depending
on the attributes exploited for prediction: collaborative filtering (CF) and
content-based (CB) methods. CF models use only the user-item rating matrix for
predictions. In our setting, this is the user-course grade matrix with grades for
each user in all courses he or she has completed. CB methods also have access to
the rating matrix but additionally incorporate either user profiles, item
descriptions, or both. For grade prediction, these features consist of student
demographics, performance history, and course features. These are collectively
referred to as \textit{content features}.

The present effort explores a variety of recommender system methods that
leverage content features, blending CF and CB modeling ideas in powerful hybrid
models. We also add data from students who transfer to the university and use
various feature importance metrics to explore the different student
characteristics with regards to course performance. This final analysis yields
several interesting insights regarding the importance of instructors and the
differing factors that affect transfer and non-transfer (native) student
performance.

The task of predicting students' grades in the coming term is complicated by the
ever-expanding volume of data from increasing student enrollments and by the
continually shifting characteristics of the overall student population.
Furthermore, recommender systems typically have issues predicting for previously
unseen users and items; these issues are collectively called the
\textit{cold-start problem}. Our experiments show that Factorization Machines
(FMs), Random Forests (RFs), and the Personalized Multi-Linear Regression (PMLR)
model achieve the lowest prediction error. While these methods also struggle
with the cold-start problem, we found that incorporation of key content features
results in improved predictions from at least one of the top three methods for
each cold-start scenario. Application of a novel feature selection metric for
the FM model greatly improves its accuracy. With the improved FM model, we were
also able to devise a hybrid FM-RF method that outperforms all three individual
methods, exploiting the strengths of both models to overcome the cold-start
problem.

\section{Literature Review}

Recommendation methods include include nearest neighbor approaches
\cite{linden_amazon.com_2003} \cite{adamic_search_2001} \cite{huete_using_2012},
matrix factorization (MF) for collaborative filtering \cite{koren_advances_2011}
\cite{shan_generalized_2010}, restricted boltzmann machines
\cite{salakhutdinov_restricted_2007}, and topic modeling methods
\cite{alsumait_embedding_2010}, among others. Much of the research pertaining to
recommendation tasks has been conducted in the domain of e-commerce. In
particular, the task of movie recommendation was widely studied during the
Netflix Prize competition \cite{bell_bellkor_2007}. Many companies have come to
rely heavily on large-scale recommender systems for reducing information
overload and targeting advertisements in industrial settings
\cite{linden_amazon.com_2003} \cite{kanagal_supercharging_2012}
\cite{shani_mdp-based_2005}.

In this paper we study the application of recommender system technology to
student grade prediction within a traditional university setting. Similar work
deals mostly with online learning environments, such as Massive Open Online
Courses (MOOCs) \cite{romero_data_2008} \cite{romero_predicting_2013}
\cite{pena-ayala_educational_2014} and Intelligent Tutoring Systems (ITS)
\cite{thai-nghe_matrix_2011} \cite{thai-nghe_multi-relational_2011}
\cite{nedungadi_personalized_2016} \cite{thai-nghe_using_2012}. In contrast, the
goal in the present study is to predict grades in a traditional university
learning environment. Online learning environments provide a variety of detailed
student behavior data that is not available in traditional learning
environments. The timeline of those studies is also more granular. In MOOCs,
studies on predicting student performance (PSP) usually seek to predict grades
for homeworks, quizzes, and exams within a single course. Some studies do span
multiple courses, but the timelines are often a single year or less
\cite{pena-ayala_educational_2014}.

Despite the differences, there are some interesting similarities between the
present work and recent work in predicting correct first attempt (CFA) in ITS.
CFA is a binary indicator: 0 if a student gets the problem/task correct on the
first try and 1 otherwise. These binary targets are aggregated to produce
real-valued "ratings" in the $[0,1]$ range. Thai-nghe utilizes standard MF and
slightly customized tensor factorization (TF) models
\cite{thai-nghe_matrix_2011}. Results demonstrate that MF and TF methods achieve
good predictive performance when compared with logistic regression and simple
baselines. The MF models used in that study are subsumed by the FM model used in
the present study. The TF model was employed to capture 3-way user-item-time
interactions. While we do explore the utility of 1- and 2-way interaction terms,
we do not explore 3-way interactions in this study. This would be an interesting
direction for future work.

Both \cite{thai-nghe_multi-relational_2011} and
\cite{nedungadi_personalized_2016} apply multi-relational matrix factorization
(MRMF) models to CFA prediction. This method simultaneously performs MF on
several "rating" matrices for complementary relations, such as "student solves a
problem", "problem requires a skill", and "student has a skill." Jointly
decomposing several matrices into partially shared factorized matrices is
empirically shown to more effectively regularize the entities in the primary
relation, yielding improved predictions. Applying unequal weights (WMRMF) to
these joint learning objectives can further improve results. Adding bias terms
to both the MRMF and WMRMF models can achieve even further improvements.
Thia-Nghe et al. also apply the FM model to the task of CFA prediction
\cite{thai-nghe_using_2012}. However, they limit their data to the user-item
matrix and learn the model with SGD. In this setting, the FM model reduces to a
regularized Singular Value Decomposition (SVD) with bias terms and a global
intercept. We studied the ability of this and other similar models for
predicting grades on a subset of the present dataset in a prior work
\cite{sweeney_next-term_2015}.

ITS researchers have also been very interested in student modelling and
knowledge tracing. In a seminal paper, Corbett and Anderson presented an
analysis whose task was modeling the changing knowledge states of students
during skill acquisition \cite{corbett_knowledge_1994}. These student models
often form core components of ITS systems. Desmarais and Baker provide a broad
review of ITS systems in \cite{desmarais_review_2011}. One example of recent
state-of-the-art includes Xu and Mostow's logistic regression dynamic bayes net
(LR-DBN) model for tracing multiple student subskills over time
\cite{xu_comparison_2012}. They also recently conducted a morphological analysis
of prominent student models, surveying recent methods and identifying gaps to be
filled \cite{xu_unified_2015}. Barnes and Stamper demonstrated another
interesting direction with their use of Markov Decision Processes (MDPs) for
automatically generating hints for an intelligent logic proof tutor
\cite{barnes_toward_2008}. Baker et. al also developed systems that can adapt
and intervene based on predictions of future student performance, and they were
able to show these interventions were effective in improving students'
experiences \cite{baker2008improving}. It would be very interesting to explore
the modeling techniques leveraged in these studies for the task of grade
prediction in future work. The present study makes no attempt to model learner
knowledge states or to provide interactive feedback during activities.

While most recent research is focused on online learning, there are some recent
studies which perform grade prediction in university settings as we do. In one
such study, comments from learning reflection assignments in a course are used
as features for grade prediction \cite{sorour_predictive_2015}. Latent Semantic
Analysis (LSA) is used to extract topics and reduce dimensionality of the
comments. Then k-means is used to cluster the comments into per-grade clusters.
Rhodes et. al. also leverage lexical features from student self-reflection for
performance prediction \cite{rhodes_using_2013}. By analyzing these features
together with student effort features (time spent writing solutions), they find
that self-explanation vocabulary correlates with amount of effort expended. The
features in these studies consist only of bag-of-words term vectors. The present
study does not leverage any such text features, since none were readily
available in our dataset.

In our previous work, we studied the next-term grade prediction task without
incorporating content features \cite{sweeney_next-term_2015} (only CF methods).
The present effort expands upon this work to explore a variety of methods that
leverage content features, blending CF and CB modeling ideas in powerful hybrid
models. In another recent study by Elbadrawy et. al.
\cite{elbadrawy_personalized_2015}, the authors devise a custom mixed-membership
multi-linear regression model (PMLR -- mentioned above) to predict student
grades in a traditional university setting. They use a variety of data,
including grades and learning management system (LMS) features. Use of LMS data
allows prediction of grades at the "activity" level -- individual assessments
within a given course. The present study does not use data from or predict
grades for individual activities. Instead, grades are predicted only at the
course level using only past student grades. We do, however, compare against the
PMLR model in our study and find it to be one of the top three performers.

\section{Problem Formulation}

Given a database of (student, course) dyads (i.e., pairs) with associated
content features for the course, student, and course instructor, our goal is to
predict grades for each student for the next enrollment term. More formally, we
have $n$ students and $m$ courses, comprising an $n \times m$ sparse grade
matrix $G$, where
$\left\{G_{ij} \in \mathbb{R}\ \middle|\ 0 \le G_{ij} \le 4\right\}$
is the grade student $i$ earned in course $j$. This is the primary source of
information leveraged by matrix factorization techniques, such as Singular Value
Decomposition (SVD). For those methods, the task is often cast as a matrix
completion problem. However, we do not wish to complete the entire matrix. We
assume domain knowledge will be available to filter the list of empty cells to
only those relevant to a particuar student. We are further restricted in our
testing efforts to only those courses students select themselves, rather than
all those which they could select from.

We consider each cell to be a (student, course) dyad and represent it as a
feature vector $\bm{X}_{ij} \in \mathbb{R}^{1 \times p}$. The first $n$
variables are a 1-of-$n$ vector representing the one-hot-encoded student IDs and
the next $m$ are a 1-of-$m$ vector for the course IDs. The rest of the $p$ total
features consist of content features generated by student $i$ taking course $j$.
We train our models on all feature vectors $\bm{X}_{ij}$ preceding the current
term and predict grades $\hat{G}_{ij}$ for all feature vectors $\bm{X}_{ij}$ in
the current term. This setup is critical for avoiding data leakage -- a severe
experimental misstep that can inflate real-world significance of performance
metrics \cite{pardos_real_2012}. For fair and effective evaluation of the
proposed methods, we train one model per academic term in the dataset, and we
use that model to predict only for this "current term." All results presented in
this paper represent an aggregate of predictions obtained with this sequential
train-predict loop.

\section{Dataset Description}

The data used for this study comes from a public university~\footnote{George
Mason University (GMU)} with an enrollment of ~33,000 students as of Fall 2014
\cite{gmu_facts_figures_2014}. Observations begin in Summer 2009 and continue
until Spring 2014, for a total of 15 terms. All students whose cohorts pre-date
this time period were excluded from the data. After preprocessing, there are
30,754 students declared in one of 144 majors, each of which belongs to one of
13 colleges. 584,179 (65.29\%) of these are transfer students and the other
310,557 are non-transfer (34.71\%). During this time period, these students have
taken 9,085 unique courses, each of which is classified as one of 161
disciplines and taught by one of 6,347 instructors. After discarding records
with no grades or grades which do not translate to the A-F scale (such as
withdrawals and audits), we have 894,736 dyads. 584,179 (65.29\%) of these are
transfer students and 310,557 (34.71\%) are non-transfer (native) students. All
this data was collected and anonymized in accordance with Institutional Review
Board (IRB) policies.

The A-F letter grades have nominal equivalents in the range 0-4, so the target
space is actually discrete. As a result, this problem could be cast as either
classification or regression. While some classification methods were explored in
preliminary experiments, these all failed to capture the ordinal nature of the
data. In order to properly capture this ordering information, we cast the
problem as a regression task.

\subsection{Content Features}

For each dyad, we have a variety of student, course, and instructor features,
either categorical or real-valued. For each of these three entities, we have
unique identifiers and a variety of other information. For students, we have
demographics data, such as age, race, sex, high school CEEB code and GPA, zip
code, and 1600-scale SAT scores. For each dyad, we have the declared major of
the student and the grade earned. For each term, we have the GPA from the
previous term as well as the cumulative GPA. We have the number of credit hours
the student is enrolled in during the current term and the number of credit
hours attempted up to the current term. We also have an academic level obtained
from credit hours attempted. Finally, we annotate each term for each student
with that student's relative term number. This feature reflects the number of
terms the student has taken courses for.

There is also a variety of data for courses. Each course belongs to a particular
discipline, is worth a fixed number of credit hours, and is assigned a
particular course level. For each term, we have the aggregate GPA of the course
from the previous term as well as the cumulative aggregate GPA over all terms
the course has previously been offerred (in our dataset). We have the number of
students enrolled in all sections during this term, as well as the total number
of students enrolled for all prior terms the course has been offered. In
addition to information specific to the course itself, the course side
information also includes instructor information. For each instructor, we have
his or her classification (Adjunct, Full time, Part time, Graduate Resarch
Assistant, Graduate Teaching Assistant), rank (Instructor, Assistant Professor,
Associate Professor, Eminent Scholar, University Professor), and tenure status
(Term, Tenure-track, Tenured). Transfer course records are mapped to GMU
equivalents. Instructor information for these records is mostly absent. So we
use the ID of the institution of origin to substitute for an instructor ID and
leave out the other instructor features. A more detailed listing of the features
used in this study is given in \nameref{feature-descriptions}.

\subsection{Cold Start Predictions}

In the context of the next-term prediction task, cold-start records are defined
as (student, course) dyads for which either or both the student and course
appear in a term (prediction phase) but do not appear in any of the previous
terms (training phase). In our dataset, 447,378 (50.00\%) dyads are
non-cold-start and 447,358 (50.00\%) are cold-start. 41,843 (9.35\%) of these
are dyads for which both the student and the course are cold-start. 389,449
(87.06\%) are student-only cold-start, and 16,066 (3.59\%) are course-only
cold-start. Table \ref{tab:cs-prop-by-term} breaks down the proportion of
cold-start records by academic term. For instance, 98.69\% of the dyads are
cold-start for the Fall 2009 cohort. The only previous enrollment was in Summer
2009, so we know the other 1.31\% of the dyads represent students who enrolled
in the previous summer.

\begin{table}[tb]
    \centering
    \caption{Cold Start Proportion By Academic Term} \label{tab:cs-prop-by-term}
    \begin{tabular}[tb]{ l r r r r r }
        \toprule
          Term        & Term \#  &  Dyads    &  NCS     &  CS      &  \% CS   \\
        \midrule
          '09 Summer  &   0      &      252  &       0  &     252  &  100.00  \\
          '09 Fall    &   1      &   62,074  &     813  &  61,261  &   98.69  \\
          '10 Spring  &   2      &   41,075  &  17,545  &  23,530  &   57.29  \\
          '10 Summer  &   3      &    3,577  &   3,165  &     412  &   11.52  \\
          '10 Fall    &   4      &   80,287  &  21,685  &  58,602  &   72.99  \\
          '11 Spring  &   5      &   64,295  &  37,358  &  26,937  &   41.90  \\
          '11 Summer  &   6      &    7,043  &   6,734  &     309  &    4.39  \\
          '11 Fall    &   7      &   98,673  &  36,166  &  62,507  &   63.35  \\
          '12 Spring  &   8      &   80,200  &  53,587  &  26,613  &   33.18  \\
          '12 Summer  &   9      &    8,880  &   8,546  &     334  &    3.76  \\
          '12 Fall    &  10      &  107,511  &  51,130  &  56,381  &   52.44  \\
          '13 Spring  &  11      &  100,342  &  64,720  &  35,622  &   35.50  \\
          '13 Summer  &  12      &   10,678  &  10,388  &     290  &    2.72  \\
          '13 Fall    &  13      &  125,312  &  58,901  &  66,411  &   53.00  \\
          '14 Spring  &  14      &  104,537  &  76,640  &  27,897  &   26.69  \\
        \bottomrule
    \end{tabular}
    \rule{0pt}{3ex}\\
    {\footnotesize
     CS: Cold-Start, NCS: Non Cold-Start\\
     Term \# denotes chronological ordering of academic terms in dataset.
     Cold-start dyads have either a new student, a new course, or both.}
\end{table}

\section{Methods}

We explore three classes of methods for the next-term student grade prediction
task. These are (1) simple baselines, (2) MF-based methods, and (3) common
regression models. The first two classes are methods which do not incorporate
content features. The last consists of traditional regression methods which must
necessarily incorporate content features. Since the FM model can use any
features, it falls into (3) when incorporating content features and (2)
otherwise.

\subsection{Simple Baselines}

We devised three simple baselines to better understand the effect of leveraging
three types of central tendencies in the data.

\begin{itemize}
    \itemsep0em
    \item  \textit{Uniform Random (UR)}: Randomly predict grades from a uniform
        distribution over the range [0, 4].
    \item  \textit{Global Mean (GM)}: Predict grades using the mean of all
        previously observed grades.
    \item  \textit{Mean of Means (MoM)}: Predict grades using an average of the
        global mean, the per-student mean, and the per-course mean.
\end{itemize}

The UR method illustrates the result of making predictions by randomly guessing.
The GM method illustrates the informative value of the overall central tendency
of the data, often called the \textit{global intercept}. The MoM method takes
this strategy one step further, also incorporating the per-student and
per-course (row and column) averages. When compared to the GM method, this
illustrates the added benefit of a small level of personalization and historical
knowledge about the course difficulty. In the cold-start setting, we allow the
MoM method to use whatever information is available. For a particular cold-start
dyad, either the student or course may be present, but not both. If neither is
present, it reduces to the GM method.

\subsection{Matrix Factorization Methods}

We look at three methods based on Matrix Factorization (MF):

\begin{itemize}
    \itemsep0em
    \item  \textit{Singular Value Decomposition (SVD)}.
    \item  \textit{SVD-kNN}: SVD post-processed with kNN~\cite{arkadiusz_paterek_improving_2007}.
    \item  \textit{Factorization Machine (FM)}~\cite{rendle_factorization_2012}.
\end{itemize}

Each of these models attempt to capture the pairwise interaction of features by
decomposing the feature space into a $k$-rank reduced subspace. This results in
two sets of latent vectors: one for the courses ($v_j \in \mathbb{R}^k$), and
one for the students ($v_i \in \mathbb{R}^k$), which can be thought of as latent
course characteristics and latent student competencies or knowledge states for each
of these general course characteristics, respectively. These latent feature
vectors can be considered a less noisy, condensed representation of the student
and course information. For SVD, each grade is simply predicted as the dot
product of the latent student and course feature vectors. We call this the
factorized 2-way interaction of the student with the course. This differs from a
simple linear regression in its use of latent feature vectors to deal with
sparsity in the data. It also differs in its use of 2-way interactions, which
capture the effects that interactions of two predictor variables have on the
target variable (the grade).
\begin{equation}
    \hat{G_{ij}} = \sum_{f=1}^k v_{i,f} v_{j,f} = v_i^T v_j.
\end{equation}
Paterek \cite{arkadiusz_paterek_improving_2007} showed that post-processing SVD
with k-nearest neighbors (kNN) yields improved predictive performance. In
essence, the predicted grade for a course is replaced with the predicted grade
for the most similar course. Similarity is calculated as cosine similarity in
the latent feature space. This has been shown to be effective in practice.

Pure collaborative filtering (CF) methods such as SVD and SVD-kNN are unable to
make predictions for cold-start records. Without any previous observations for a
student or a course, no latent student competencies or latent course
characteristics can be learned. Content-based (CB) methods handle course
cold-start by incorporating features describing each item and/or student
profiles such as demographics \cite{adomavicius_toward_2005}. FMs can
incorporate arbitrary content features while also leveraging the sparse
student-course grade matrix. This capability means the FM is a hybrid model
which can incorporate the data traditionally used by CB and CF methods in one
model. Combined with suitable feature engineering, this allows FMs to subsume
most state-of-the-art factorization-based recommender system models developed up
to this point \cite{rendle_factorization_2012}.

We observe that the FM model is able to capture all of the information captured by
the simple baselines as well as the information captured by SVD. In general, it
captures the global central tendency, 1-way (linear) relationships between the
predictors and the grade (bias terms), and 2-way factorized interactions between
each predictor and the grade:
\begin{align}
    \hat{G_{ij}} & = \overbrace{w_0}^\text{global intercept} +
                     \overbrace{
                         \sum_{l=1}^p w_l x_l
                     }^\text{1-way interactions} +
                     \overbrace{
                         \sum_{l=1}^p \sum_{l'=l+1}^p x_l x_l' (v_l^T v_{l'})
                     }^\text{2-way factorized interactions}
\end{align}
We compare SVD and SVD-kNN to the FM model trained using only the student-course
grade matrix $G$. With this data, the model reduces to the sum of global
($w_0$), student ($w_i$), and course ($w_j$) bias terms and the factorized
interaction of the student with the course ($v_i^T v_j$). This last term is the
same dot product seen in the SVD model.
\begin{align}
      \hat{G_{ij}} & = w_0 + w_i + w_j + v_i^T v_j.
\end{align}
We use the fastFM library~\cite{bayer_fastfm_2015} for a fast implementation of
the FM algorithm. The model parameters are learned with Gibbs sampling.

\subsection{Common Regression Models}

We tested four different regression models:
\begin{itemize}
    \itemsep0em
    \item  \textit{Random Forest (RF)}
    \item  \textit{Stochastic Gradient Descent (SGD) Regression}
    \item  \textit{k-Nearest Neighbors (kNN)}
    \item  \textit{Personalized Multi-Linear Regression (PMLR)}
\end{itemize}

We leveraged the scikit-learn library \cite{scikit-learn} for the first three,
which are common regression models. Parameter settings for each method were
found using grid search on a sampled held-out set. These are listed in
\nameref{parameter-settings}. We briefly describe each of these models.

\subsubsection{Random Forest}

The Random Forest algorithm combines a group of random decision trees in a
bagging ensemble \cite{breiman_random_2001}. A single non-random decision tree
is constructed by discovering the most informative questions to ask in order to
split all samples into groups with similar target attribute values. Each
question splits the data into two or more groups by thresholding some feature of
the samples. So we end up with a tree of decision nodes. For regression, the
most informative questions are those that produce leaf nodes whose mean squared
error (MSE) are minimal among all possible splits.

Tree construction stops once an additional split would not reduce MSE. Since
this usually overfits, an early-termination criterion is often specified. This
is usually maximum tree depth or minimum number of nodes at each leaf; we used
the former for our experiments. Once built, the tree can be used for regression
of new data samples. A sample is run through the decision-making sequence
defined by the structure of the tree until reaching a leaf. Then the prediction
is the mean of the grades of the samples at that node when training finished.

A random decision tree results from learning a decision tree on a bootstrap data
sample and considering a random subset of features for each split.  This
``feature bagging'' is done to reduce correlation between the trees. The Random
Forest then combines many of these trees in a weighted averaging approach to
make decisions regarding unseen data. This reduces the variance of the
individual trees while retaining the low-bias.

\subsubsection{Stochastic Gradient Descent (SGD) Regression}

The SGD regression method learns a least squares linear regression fit under an
L1 regularization penalty (absolute norm). The least squares fit minimizes the
squared difference between actual and predicted grades, while the L1
regularization penalty encourages feature sparsity. In particular, unimportant
parameters have a tendency to be pushed towards 0, so the L1 penalty operates as
a kind of online feature selection. In contrast to ordinary least squares (OLS),
this method is an approximate best fit, learned using stochastic gradient
descent (SGD). SGD is a gradient-based optimization technique that updates the
model parameters incrementally, rather than on the entire training set at once
(which is what normal gradient descent would do). This reduces overfitting and
significantly improves training time.

\subsubsection{k-Nearest Neighbors}

The $k$-Nearest Neighbors (kNN) algorithm is a classic method for clustering
samples based on similarity. These clusterings can be used for regression in the
following manner. First a pairwise distance metric is used to identify the $k$
most similar neighbors among all dyads in the training set. The grade is then
predicted using local interpolation of the targets associated with these
neighbors. Many different distance metrics can be used; for our experiments, we
use standard Euclidean distance:
\begin{align}
    Euclidean(\bm{X}_{i,j}, \bm{X}_{i', j'}) =
        \sqrt{\sum_{f=1}^p \left(
            \bm{X}_{i,j,f} - \bm{X}_{i',j',f}
        \right)^2}.
\end{align}
To predict a grade for a new dyad $(i, j)$, the Euclidean distance from $(i, j)$
to every dyad in the training set is computed. The $k$ dyads $(i', j')$ with the
smallest distance are selected and placed into a set of neighbors
$\mathcal{N}_{i,j}$. The grade for student $i$ in course $j$ is then predicted
as the uniformly weighted average of these neighbors' grades.

\subsubsection{Personalized Multi-Linear Regression}

We also employ the more recently developed Personalized Multi-Linear Regression
model from \cite{elbadrawy_personalized_2015}. The original model predicts a
missing grade $G_{ij}$ for student $i$ in course $j$ using:
\begin{align}
    \hat{G}_{ij} &= s_i + c_j + \sum_{l=1}^k P_{il}
                                \sum_{f=1}^p W_{lf} \bm{X}_{ijf}  \\
                 &= s_i + c_j + P_i W \bm{X}_{ij},
\end{align}
where $s_i$ is a bias term for student $i$, $c_j$ is a bias term for course $j$.
PMLR uses a linear combination of $k$ regression models weighted on a
per-student basis. $P_i$ is the $1 \times k$ vector of model weights for student
$i$ and $W$ is the $k \times p$ matrix of regression coefficients. $\bm{X}_{ij}$ is
the feature vector generated by student $i$ taking course $j$. The model is
learned using the following objective function. Root Mean Squared Error (RMSE)
is used as the loss function $\mathcal{L}(\cdot)$, $P$, and $W$ are regularized
using the squared Frobenius norm, and all parameters are constrained to be
non-negative.
\begin{align}
    &\underset{P, W, s, c}{\text{minimize }}
        \mathcal{L}(P, W, s, c)
            + \lambda_W (\norm{P}_F^2 + \norm{W}_F^2)
\end{align}

We implemented our own version of the PMLR model, adding a global intercept term
$w_0$ and a regularization on the bias terms controlled by parameter
$\lambda_B$. The resulting model and objective function are shown below. All
parameters are contrained to be non-negative, as in the original model.
\begin{align}
    &\hat{G}_{ij} =  \overbrace{w_0}^\text{new} + s_i + c_j + P_i W \bm{X}_{ij} \\
    &\underset{P, W, s, c}{\text{minimize }}
        \mathcal{L}(P, W, s, c)
            + \lambda_W (\norm{P}_F^2 + \norm{W}_F^2)
            + \underbrace{\lambda_B (\norm{s}_F^2 + \norm{c}_F^2)}_\text{new}
\end{align}

\section{Experimental Results and Discussion}

\subsection{Metrics}

Evaluations are performed in terms of two common regression metrics: Root Mean
Squared Error (RMSE) and Mean Absolute Error (MAE).
\begin{align*}
    RMSE = \sqrt{ \frac{\sum_{i=1}^N (\hat{G}_{ij} - G_{ij})^2 }{N} }
    \hspace{15mm}
    MAE = \frac{\sum_{i=1}^N | \hat{G}_{ij} - G_{ij} | }{N}
\end{align*}
RMSE penalizes severe prediction errors more heavily than small ones. Given our
task, we would prefer not to declare a method the best if it performs very well
for half the students and very poorly for the other half. Hence RMSE is the
metric we use to compare methods. MAE allows us to understand the range of
grades we might actually be predicting. If a student is actually going to get a
B, and the MAE is 0.33, we expect our model to predict either a B-, B, or B+.
All MAE measurements are accompanied by the standard deviation of the absolute
errors.

\subsection{Feature Preprocessing and Selection}

For preprocessing, missing values for each real-valued feature were filled in
using the medians. This was not done for the FM model, since it handles missing
data without loss of performance and performs worse with the median-value
imputation. After this step, the real-valued attributes were scaled using
Z-score scaling. Finally, the predictions of all methods are post-processed to
bound the grade predictions to the range [0, 4].

Unlike any of the other models used, FMs are capable of learning effectively
from both categorical and real-valued features. We would like to maximize the
amount of differentiating information which can be captured by the 2-way
interaction factors of the model. For this purpose, when we have the choice
between encoding a feature as categorical (one-hot encoded) or real-valued
(single feature with ordinal value), we choose the categorical encoding. This
allows unique 2-way interactions to be learned for each combination of
categories for all categorical features. Among the techniques which can
incorporate content features, the FM model is the only one which leverages MF
techniques to learn these sparse 2-way interactions effectively. With this in
mind, we do not include highly sparse features (such as the instructor IDs) as
training data for the other models. This reduces computational overhead at an
insignificant loss to performance.

\subsubsection{FM Feature Selection: A Novel Importance Metric}

In our experiments, we found that the FM model has particularly poor feature
selection capabilities, unlike the L-1-regularized regression model and the
decision-tree-based models. We introduce a new feature importance metric called
Mean Absolute Deviation Importance (MADImp).

Inspired by the work of \cite{elbadrawy_personalized_2015}, MADImp is a
method for computing the importance of each feature in any generalized linear
model (GLM). Elbadrawy et. al. compute importance for a model term as the
proportion of the overall prediction accounted for by that term, normalized over
all records. This is greatly simplified by the non-negativity constraints
enforced in the particular linear model used by those authors. Since not all
GLMs have non-negativity constraints, we cannot use simple averaging.
Instead, we use a quantity computed from the mean absolute deviation from a
global intercept term.

For each dyad, the active features (non-zero) are involved in one or more
additive interaction terms which move the prediction away from the global
intercept $w_0$. To measure importance, we sum the absolute values of these
additive terms for a particular feature and divide by the total absolute
deviations caused by all features. Importance is measured as the proportion
of the absolute deviation from $w_0$ accounted for by a particular feature. We
average this over all records -- hence the name Mean Absolute Deviation
Importance (MADImp). For a practical example that solidifes these ideas, see
\nameref{madimp-example}.

We now define MADImp more formally. Throughout this section we use $d$ as
shorthand for some dyad indices $i, j$. The importance of each feature for a
single dyad is calculated as:
\begin{equation} \label{eq:importance-single-dyad}
    I(\bm{X}_{d,f}) =
        \frac{\sigma_1(\bm{X}_{d,f}) + \sigma_2(\bm{X}_{d,f})}{T_d},
\end{equation}
where $\sigma_1(\bm{X}_{d,f})$ is the deviation from 1-way interactions of
feature $f$ in dyad $d$, $\sigma_2(\bm{X}_{d,f})$ is the deviation from 2-way
interactions of feature $f$ in dyad $d$, and $T_d$ is the total deviation from
the global intercept $w_0$ in the estimation for dyad $d$.
\begin{align}
    \sigma_1(\bm{X}_{d,f})
        &= |w_f \bm{X}_{d,f}|  \\
    \sigma_2(\bm{X}_{d,f})
        &= \bm{X}_{d,f}^2 \sum_{f'=1, f' \neq f}^p
            \frac{|\bm{X}_{d,f} Z_{f,f'}|}
                 {|\bm{X}_{d,f}| + |\bm{X}_{d,f'}|}  \\
    T_d
        &= \sum_{f=1}^p |w_f \bm{X}_{d,f}| +
            \sum_{f=1}^p \sum_{f'=f+1}^p
                |\bm{X}_{d,f} \bm{X}_{d,f'} Z_{f,f'}|
\end{align}

The sum of the importance measures of all features for a dyad equals the total
deviation for that dyad: $\sum_{f=1}^p I(\bm{X}_{d,f}) = T_d$. We can calculate
the overall importance of a feature in our training dataset through:
\begin{equation}
    I(\bm{X}_{*,f}) = \sum_{d=1}^p I(\bm{X}_{d,f})
\end{equation}
In general, to calculate MADImp for $\bm{X}_{*,f}$ for any GLM, group all terms
$\bm{X}_{*,f}$ is involved in and take the absolute value. Do the same for all
other $\bm{X}_{*,f}$. Normalize by the sum of all term deviations $\sum_{f}
\bm{X}_{*,f}$.  Finally, take the mean across all training records.

After training the FM model on all academic terms, we average these importance
metrics across terms, weighting by the number of records predicted in each term.
By applying this metric, we found that the student bias, course bias, and
instructor bias were most informative, followed by the course discipline, major,
student race, and instructor rank. The cohort, instructor class and tenure,
student term, transfer indicator, and sex were found to be less important but
still informative. Training the FM with only these features greatly improves the
results when compared with training on all features. \textbf{RMSE dropped from
1.0587 to 0.7832 -- an improvement of 26\%}. This demonstrates that using MADImp
for FM feature selection can greatly improve results and significantly reduce
the amount of effort required when searching for the most suitable features. For
the other models, we relied on their inherent feature selection capabilities.

\subsection{Grade Prediction Results}

The prediction results are broken down by non-cold-start vs. cold-start in
Tables~\ref{tab:pred-ncs-vs-cs}. We first discuss results from methods not
incorporating content features. On non-cold-start records, we see the UR and GM
methods perform worst, as expected. It is somewhat surprising to see SVD perform
worse than the MoM method. This is likely due to overfitting, which is a common
issue with SVD \cite{arkadiusz_paterek_improving_2007}. While the
post-processing is able to improve the SVD results, they are not nearly as good
as those produced by the MoM method. This indicates the per-student and
per-course biases are quite informative. The FM model outperforms all the others
by a wide margin, indicating 2-way feature interactions play an important role
in predicting performance. This also indicates that proper regularization
noticeably improves the generalization of learned patterns for prediction in
future terms.

All methods except the MoM method actually show improved results when predicting
on cold-start records. Since the GM method improves, we can conclude the grades
are simply less spread among cold-start dyads, making the task simpler than is
usually the case. This is likely an artifact of the large number of transfer
credits. These are all cold-start dyads since transfer students are all
previously unseen in the dataset. Since only passing grades can transfer from
other universities, these transfer credits shift the grade distribution
upwards and make it more centered around the mean. The MoM method goes from
being the best to being the worst. This indicates that relying on only one or
two of the bias terms leads to over-confidence in predictions.  In this setting,
the FM model is relying on essentially the same information as the MoM method,
but it is able to avoid overconfident learning patterns through the use of
Bayesian complexity control (integrating out the regularization hyperparameters)
for proper regularization.

\begin{table}[t!]
    \centering
    \caption{Non-Cold Start vs. Cold-Start Prediction RMSE}
    \label{tab:pred-ncs-vs-cs}
    \begin{tabular}{ l@{\hskip 0.15in}%
                     r@{\hskip 0.10in}c@{\hskip 0.15in}%
                     r@{\hskip 0.10in}c }
        \toprule
                       & \multicolumn{2}{c}{NCS}   &  \multicolumn{2}{c}{CS}  \\
          Method       &  RMSE    &  MAE  &  RMSE  &  MAE \\
        \midrule
          Without Content Features & & & & \\
        \midrule
          FM           & \textbf{0.7792} &  \textbf{0.56 $\pm$ 0.54}%
                       & \textbf{0.7980} &  \textbf{0.61 $\pm$ 0.51}      \\
          MoM          &  0.8643  &  $0.64  \pm  0.58$%
                       &  1.4788  &  $1.30  \pm  0.70$  \\
          SVD kNN      &  0.9249  &  $0.68  \pm  0.62$%
                       &  0.8123  &  $0.67  \pm  0.46$  \\
          SVD          &  0.9370  &  $0.69  \pm  0.63$%
                       &  0.8095  &  $0.67  \pm  0.46$  \\
          GM           &  0.9448  &  $0.72  \pm  0.61$%
                       &  0.8144  &  $0.68  \pm  0.45$  \\
          UR           &  1.8667  &  $1.54  \pm  1.06$%
                       &  1.8977  &  $1.57  \pm  1.07$  \\
        \midrule
          With Content Features & & & & \\
        \midrule
          FM           &  \textbf{0.7423} & \textbf{0.52 $\pm$ 0.53}%
                       &  0.8111  &  $0.61  \pm  0.53$  \\
          PMLR         &  0.7886  &  $0.57  \pm  0.55$%
                       &  1.0207  &  $0.75  \pm  0.70$  \\
          RF           &  0.7936  &  $0.58  \pm  0.54$%
                       &  \textbf{0.7475} & \textbf{0.59 $\pm$ 0.46}  \\
          kNN          &  0.8061  &  $0.59  \pm  0.55$%
                       &  0.8101  &  $0.65  \pm  0.49$  \\
          SGD          &  0.8207  &  $0.60  \pm  0.56$%
                       &  1.0297  &  $0.75  \pm  0.71$  \\
        \bottomrule
    \end{tabular}  \\
\end{table}

We next discuss results for prediction with content features included (see Table
\ref{tab:pred-ncs-vs-cs}). The FM model continues to produce the lowest-error
predictions. The next best method is PMLR, followed by the RF. For non-cold
start dyads, all of these methods outperform MoM, but only the FM with content
features outperforms the FM without. The same thing cannot be said for the FM on
cold-start records. In this setting, the RF model is the clear winner. The FM
model performs just slightly worse than the kNN regressor.

It is particularly interesting to notice that the FM with content features
performs worse than the FM without on cold-start prediction. This indicates the
FM model is learning patterns from prior terms that no longer hold in the
current (prediction) term. It is able to capture 2-way interactions between
features while other methods are not. So these patterns seem to be shifting
significantly as new students enroll. This problem, where the test distribution
differs significantly from the training distribution, is commonly known as
\textit{covariate shift} \cite{shimodaira_improving_2000}. Uncovering
inability to deal with covariate shift as a limiting factor in one of our best
methods allows us to target future research at overcoming this problem. We
discuss these implications in more detail in Section~\ref{discussion}

\begin{table}[t!]
    \centering
    \caption{Native vs. Transfer Prediction RMSE}
    \label{tab:pred-native-vs-transfer}
    \begin{tabular}{ l@{\hskip 0.15in}%
                     r@{\hskip 0.10in}c@{\hskip 0.15in}%
                     r@{\hskip 0.10in}c }
        \toprule
                       & \multicolumn{2}{c}{Native}   &  \multicolumn{2}{c}{Transfer}  \\
          Method       &  RMSE    &  MAE%
                       &  RMSE    &  MAE \\
        \midrule
          Without Content Features & & & & \\
        \midrule
          FM           &  \textbf{0.8054}  &  \textbf{0.58 $\pm$ 0.56}%
                       &  \textbf{0.7805}  &  \textbf{0.62 $\pm$ 0.48}  \\
          SVD-kNN      &  0.9343  &  $0.69 \pm 0.63$%
                       &  0.8345  &  $0.67 \pm 0.50$  \\
          SVD          &  0.9519  &  $0.70 \pm 0.65$%
                       &  0.8325  &  $0.67 \pm 0.49$  \\
          GM           &  0.9576  &  $0.72 \pm 0.63$%
                       &  0.8391  &  $0.68 \pm 0.49$  \\
          MoM          &  1.0076  &  $0.77 \pm 0.65$%
                       &  1.3064  &  $1.08 \pm 0.74$  \\
          UR           &  1.8679  &  $1.54 \pm 1.06$%
                       &  1.8899  &  $1.56 \pm 1.07$  \\
        \midrule
          With Content Features & & & & \\
        \midrule
          FM         &  \textbf{0.7879}  &  \textbf{0.55 $\pm$ 0.57}%
                     &  0.7807  &  \textbf{0.58 $\pm$ 0.52}  \\
          RF         &  0.8062  &  $0.59 \pm 0.55$%
                     &  \textbf{0.7515}  &  $0.59 \pm 0.47$  \\
          kNN        &  0.8346  &  $0.61 \pm 0.57$%
                     &  0.7936  &  $0.62 \pm 0.49$  \\
          PMLR       &  0.8860  &  $0.62 \pm 0.64$%
                     &  0.9258  &  $0.68 \pm 0.63$  \\
          SGD        &  0.9137  &  $0.65 \pm 0.66$%
                     &  0.9402  &  $0.69 \pm 0.64$  \\
        \bottomrule
    \end{tabular}  \\
\end{table}

Table \ref{tab:pred-native-vs-transfer} provides an alternative breakdown of the
prediction results, comparing native students to transfer students. We observe
significantly better prediction results for transfer students than for native
students and note the gap is narrowed when incorporating content features. This
observation is another reflection of the reduced spread of the transfer grade
distribution as compared to native grades. Notice that the best performing
methods closely align with the non-cold-start vs. cold-start results from Table
\ref{tab:pred-ncs-vs-cs}: FM performs best on native students while RF performs
best on transfer students. This largely reflects the proportion of cold-start
dyads from each sample: only 7.37\% of native dyads are cold-start while 42.62\%
of transfer dyads are cold-start. The greater diversity of grades among the
native students makes the content features much more important for accurate
predictions.

Table~\ref{tab:top3-cs-vs-ncs} lays out the results for the top three methods
(FM, PMLR, and RF) with a more detailed cold-start vs. non-cold-start error
breakdown. The FM model is outperformed when student information is absent; in
these cases, the RF model performs best. PMLR only performs well on
non-cold-start records. These results indicate that we can improve our next-term
grade prediction system by swapping out RF for FM whenever we are lacking prior
student information. Doing so gives an overall RMSE of 0.7443 compared to 0.7709
for FM and 0.7775 for RF. This demonstrates that such a hybrid is a viable
solution to overcoming the cold-start problems suffered by FMs.

\begin{table}[tb]
    \centering
    \caption{Cold-start vs. Non-Cold-start Error} \label{tab:top3-cs-vs-ncs}
    \begin{tabular}{ r r l c c c }
        \toprule
          Group  &  Dyad \%  &  Method  &  RMSE    &  MAE  \\
        \midrule
NCS  &  48.60  &    \textbf{FM} & \textbf{0.7423} & \textbf{0.5187 $\pm$ 0.5310}  \\
               &         &  PMLR    &  0.7890  &  0.5635 $\pm$ 0.5522  \\
               &         &  RF      &  0.7936  &  0.5837 $\pm$ 0.5377  \\
        \rule{0pt}{3ex}
CSS  &  42.31  &    \textbf{RF} & \textbf{0.7381} & \textbf{0.5867 $\pm$ 0.4478}  \\
               &         &  FM      &  0.8112  &  0.6114 $\pm$ 0.5331  \\
               &         &  PMLR    &  0.9917  &  0.7321 $\pm$ 0.6689  \\
        \rule{0pt}{3ex}
CSC  &  1.75   &    \textbf{FM} & \textbf{0.7456} & \textbf{0.5293 $\pm$ 0.5252}  \\
               &         &  RF      &  0.7776  &  0.5695 $\pm$ 0.5295  \\
               &         &  PMLR    &  1.1771  &  0.7489 $\pm$ 0.9081  \\
        \rule{0pt}{3ex}
CSB  &  4.55   &    \textbf{RF} & \textbf{0.8203} & \textbf{0.6603 $\pm$ 0.4867}  \\
               &         &  FM      &  0.8337  &  0.6614 $\pm$ 0.5075  \\
               &         &  PMLR    &  1.2060  &  0.8829 $\pm$ 0.8215  \\
        \bottomrule
    \end{tabular}
    \rule{0pt}{3ex}\\
    {\footnotesize
     CS = cold-start, NCS = non-CS, CSS = CS student-only,\\
     CSC = CS course-only, CSB = CS both}
\end{table}

\subsection{Feature Importance}

It is useful to have some notion of which features are most informative for
next-term grade prediction. This understanding can eventually help us to uncover
the relationships between student performance and the various predictor
variables available to us. These features differ between the FM, RF, and PMLR
models. While FMs can leverage sparse categorical features effectively via
matrix factorization, decision trees are typically unable to discover useful
patterns in such data. The PMLR model can leverage 1-way interactions from
categorical features, but they must rely on real-valued features to learn useful
trends in the absence of the 2-way interactions of the FM model.

\begin{figure}[tb]
    \centering
    \includegraphics[width=\textwidth]{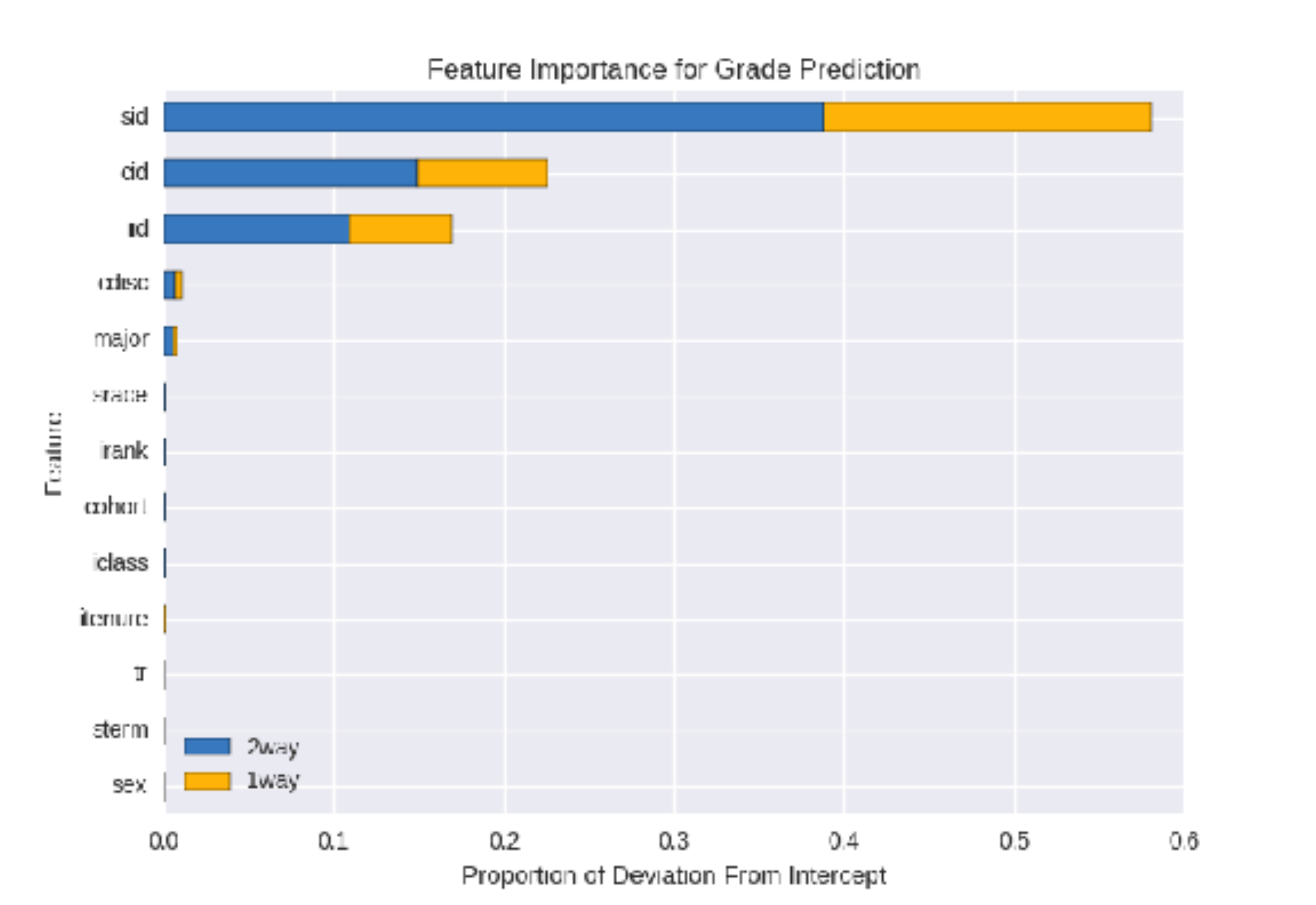}%
    \caption{FM: Overall MADImp Feature Importance}
    \label{fig:fm-fimps-all-students}%
\end{figure}

We start by examining the importance of features in the FM model.
Figure~\ref{fig:fm-fimps-all-students} shows the importance of features broken
down by 1-way and 2-way interactions. The top 3 features make up 97.3\% of the
importance. These are the student bias (sid), course bias (cid), and instructor
bias (iid). Roughly two thirds of the importance comes from 2-way interactions,
and the other third comes from the 1-way interactions. Of the non-bias terms,
the course discipline (cdisc) and major are the only features which have any
notable importance, accounting for 1.1\% and 0.8\%. This indicates that these
high-level interactions can account for the majority of variance in predictions.

\begin{figure}[tb]
    \centering
    \includegraphics[width=\textwidth]{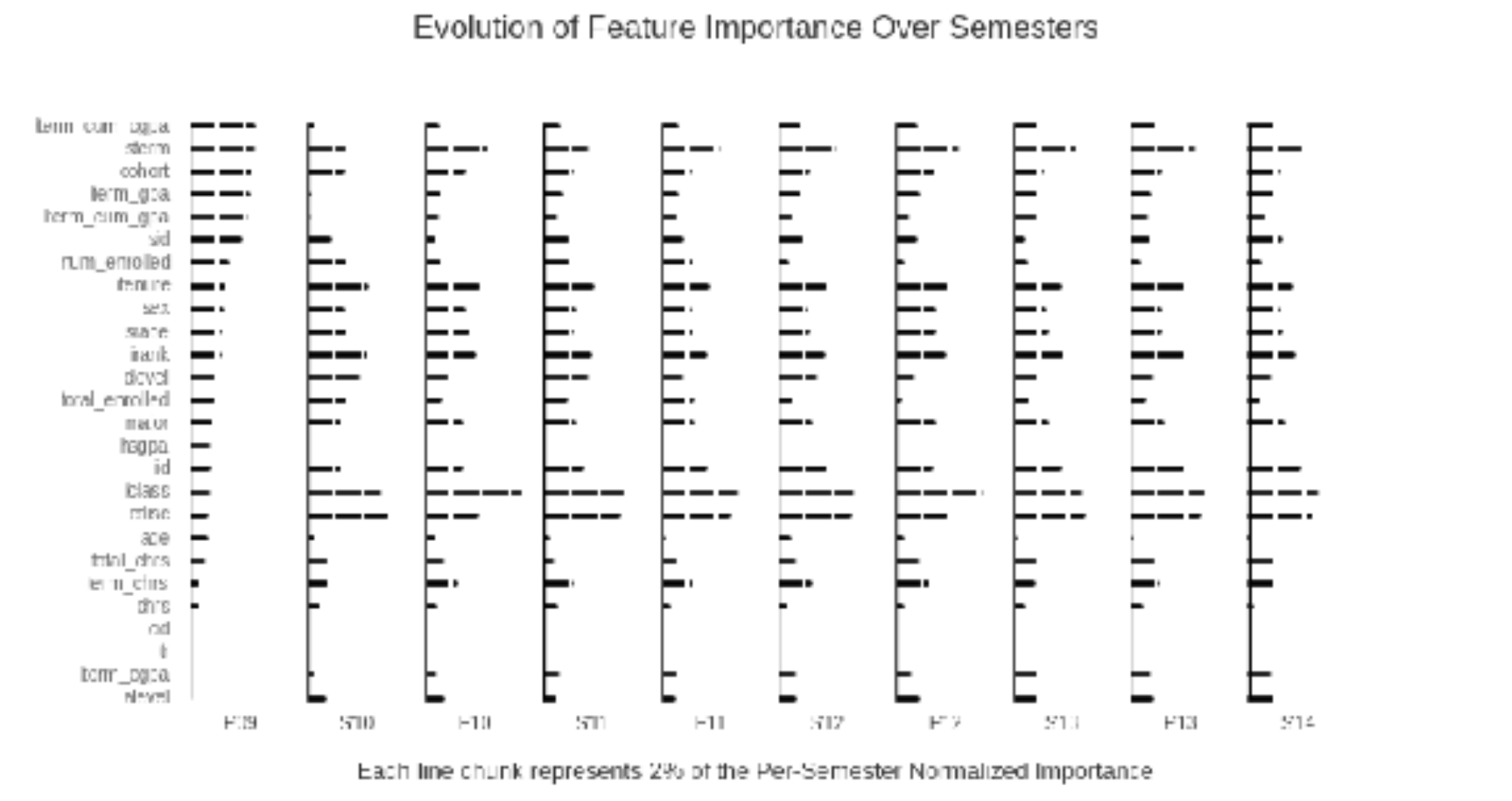}%
    \caption{PMLR: Feature Importance Evolution for All Data}
    \label{fig:pmlr-all-fimp-evol}%
\squeezeup
\end{figure}

The PMLR model does not learn 2-way interactions between features, so there is
much more variety in the importance metrics. Figure~\ref{fig:pmlr-all-fimp-evol}
shows the evolution of feature importance across terms, excluding the summer
terms (which have very few records). The individual lines represent the percent
of the overall importance accounted for by a particular feature in a particular
term. Each segment/chunk in a line represents 2\% of the overall importance. The
X-axis represents the terms, moving from Fall 2009 to Spring 2014.

We see a clear shift from the first term to the second and less drastic shifts
moving forward from there. By the last term, the instructor bias,
classification, rank, and tenure status, the number of terms a student has been
enrolled, and the course discipline emerge as the most important features. We
see that last-term GPA features gain in importance as more data is acquired. In
contrast, the number of students enrolled in a course during the current term
and in prior terms both decline in importance.

\begin{figure}[tb!]
    \centering
    \includegraphics[width=\textwidth]{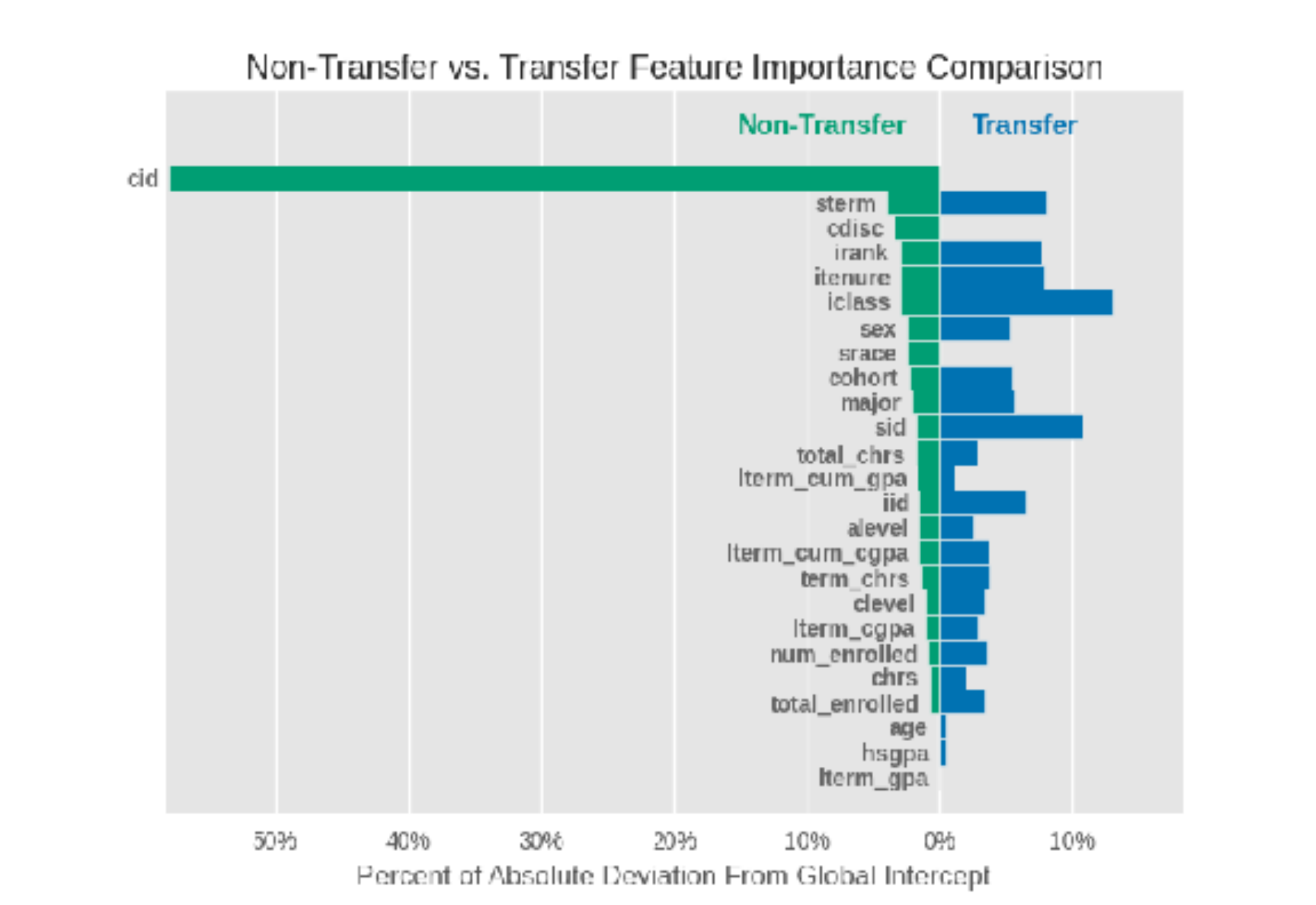}%
    \caption{PMLR: Feature Importance Comparison}
    \label{fig:pmlr-fimp-tvnt-bp}%
\end{figure}

Figure~\ref{fig:pmlr-fimp-tvnt-bp} shows the overall PMLR feature importances
for the transfer vs. the non-transfer data. These were obtained by training
separate models on the transfer data and the non-transfer data. The most
noticeable difference is the massive importance of the course bias (accounting
for nearly 60\% of overall bias) in the non-transfer data compared to the
complete lack of course bias importance in the transfer data. This is most
likely due to the mapping from transfer credits to their GMU equivalents. A
variety of courses from many different universities can be mapped to the same
GMU course. So the mapping process dilutes the usefulness of this feature. This
also explains the lack of importance of the course discipline in the transfer
data. For the transfer population, the lack of useful course bias terms causes
the PMLR model to learn larger weights for the other features. Discounting this
general trend, there are still marked differences in importance for the number
of terms a student has been enrolled, instructor classification, and the student
and instructor bias terms.

The RF model performs binary splits on the input data using one particular
feature at a time. The Gini Coefficient is used to determine the most
informative splits at each level of the tree. Gini Importance (GI) is computed
as the normalized total reduction in mean squared error (MSE) brought by each
feature over all splits. Due to the normalization, this can also be thought of
as the percent of the MSE reduction achieved. We compute GI for each term,
including predictions for cold-start records, then average them to get the final
GI proportions. The top 10 most important features of the RF in order of highest
to lowest GI are shown in Table~\ref{tab:rf-features-by-gi}.

\begin{table}[tb]
    \centering
    \caption{RF: Top 10 Features by Gini Importance (GI)}
    \label{tab:rf-features-by-gi}
    \begin{tabular}{ r l l }
      \toprule
        Feature  &  GI  &  Description  \\
      \midrule
        lterm\_gpa       &  0.4164  &  Student's last-term GPA               \\
        lterm\_cum\_cgpa &  0.2484  &  Course's last-term cumulative GPA     \\
        lterm\_cum\_gpa  &  0.0645  &  Student's last-term cumulative GPA    \\
        prior\_gpa       &  0.0539  &  Student's high school or transfer GPA \\
        lterm\_cgpa      &  0.0349  &  Course's last-term GPA                \\
        total\_chrs      &  0.0197  &  Credit hours attempted by student     \\
        term\_chrs       &  0.0175  &  Credit hours student is taking this term  \\
        total\_enrolled  &  0.0170  &  Total historical course enrollment    \\
        zip              &  0.0153  &  Student zip code                      \\
        iclass           &  0.0147  &  Instructor classification             \\
      \bottomrule
    \end{tabular}
\end{table}

For the RF model, the most informative are the grade, credit hour, and course
enrollment features. Together, these features characterize how competent a
student generally is and how difficult and well-established a course generally
is. The prior GPA gives some notion of historical competency, but this weighs in
less heavily than more recent college grades. The zip code is the one
demographic feature that shows up here, indicating that demographics are
important when no student-specific bias terms can be learned. The instructor
classification shows up again here as well.

It is particularly interesting to note that the instructor features were
important for all three methods. This indicates that the particular instructor
and his or her rank, classification, and tenure status have detectable effects
on student grades in particular courses. To our knowledge, no previous studies
have explored the effect of instructor biases or interactions for grade
prediction. Having clearly characterized their importance here, we hope further
research in this field can capitalize on the information present in such
features to improve educational data mining applications.

\subsection{Error Analysis}

\begin{figure}[H]
    \centering
        \subfloat[Error breakdown for FM predictions, excluding summer terms.]{
            \label{fig:cohort-by-tnum-err-heatmap}%
            \includegraphics[height=0.45\textheight]{%
                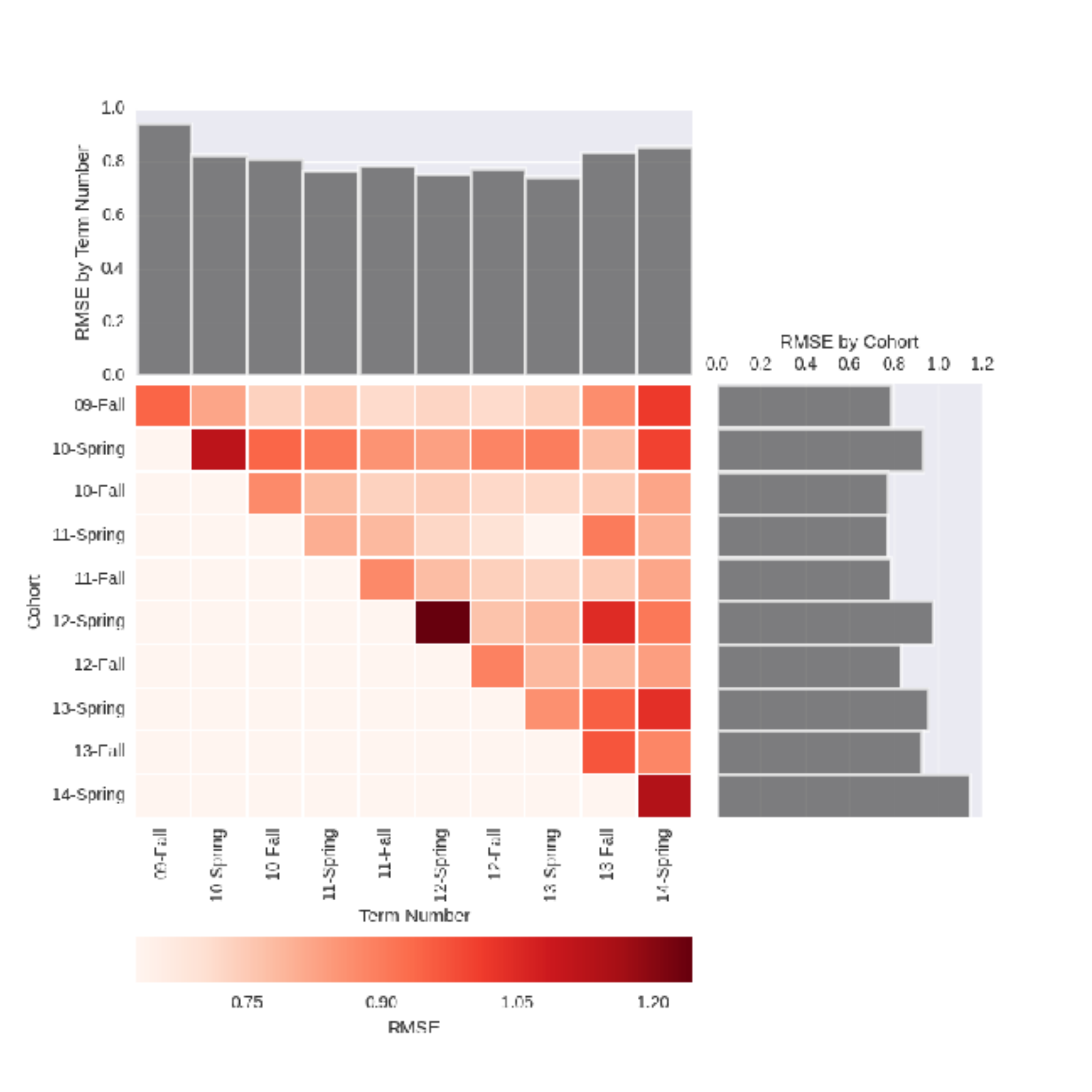}}
        \\
        \subfloat[Dyad counts for each cell in error breakdown.]{
            \label{fig:cohort-by-tnum-counts}%
            \includegraphics[height=0.45\textheight]{%
                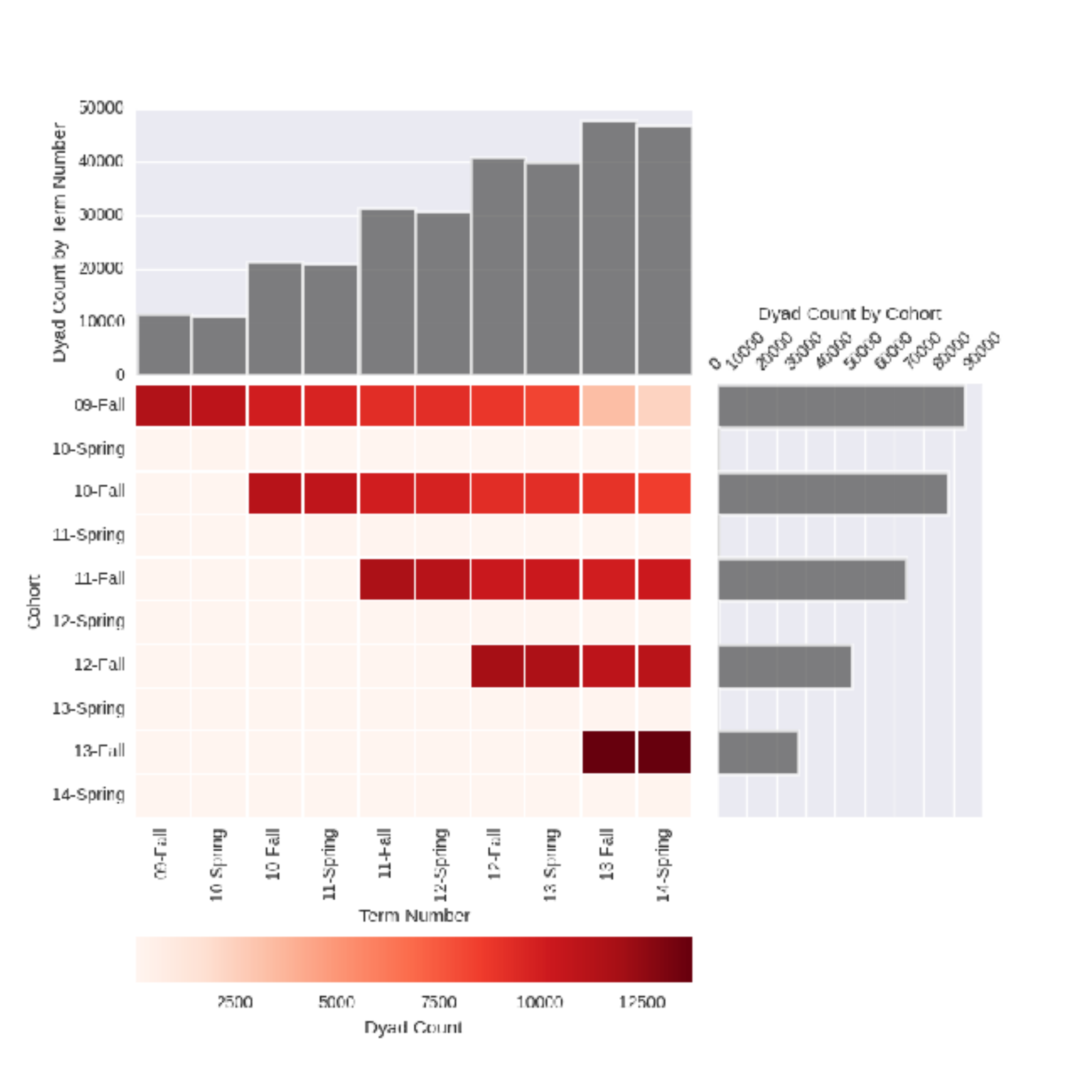}}%
    \caption{Cohort by term number error visualization for non-transfer
             students.}
    \label{fig:cohort-by-tnum-err}
\squeezeup
\end{figure}

Fig.~\ref{fig:cohort-by-tnum-err-heatmap} lays out the distribution of error for
the FM model in terms of RMSE in a cohort by term number matrix for the
non-transfer students. The cohort is the term the student was admitted to the
university, while the term number is the objective chronological ordering of the
academic terms in the dataset.

\floatsetup[figure]{style=plain,subcapbesideposition=top}
Each cell represents the aggregate RMSE for all grade predictions for the
students admitted in a particular cohort taking courses in a particular academic
term. The bar chart on top of the heatmap shows the per-term RMSE, and the bar
chart on the right shows the per-cohort RMSE. These errors are from FM with the
best settings using content features and predicting for all records, including
cold-start.

Note that we have excluded Summer terms, which account for only 2.67\% of the
predictions. This clarifies the trends observed. Also note that Spring cohorts
account for only 1.10\% of the total predictions after removing summer terms.
Hence they have much more variability than the Fall terms.
Fig.~\ref{fig:cohort-by-tnum-counts} shows a heatmap of the dyad counts for each
of the cohort/term cells.

Looking down the diagonal, we see that predictions for each cohort for the first
term are generally poorer than predictions for subsequent terms. This reflects
the increased difficulty of cold-start predictions, since these are all
cold-start students and may also be cold-start courses (new courses). Following
each row from left to right, we expect the predictions to decrease in error as
we accumulate more information about the students in this cohort. With the
exception of the Fall of 2013 and Spring of 2014, we do see this trend. So in
general, our predictions improve as we accumulate more historical grade data.

\section{Discussion} \label{discussion}

\subsection{Predictive Performance}

We found that a FM-RF hybrid is an effective method of overcoming the
cold-start limitations of FMs and is the most effective method in general.
However, we also note that more sophisticated methods have been developed to
improve performance of MF-based methods on cold-start prediction tasks. One
notable example is the attribute-to-latent-feature mapping developed by
\cite{gantner_learning_2010}. Such an approach might produce better results
and/or be computationally cheaper than the FM-RF hybrid.

We also uncovered covariate shift as a likely source of error when using FMs for
cold-start prediction. When the test distribution differs from the training
distribution, the FM model is liable to learn overconfident 2-way interactions
that reduce its ability to generalize. The RF model does not seem to suffer from
this problem, but it is also clearly unable to capture 2-way interactions.
Discovering new ways to handle covariate shift while learning 2-way interactions
might be another viable way to overcome cold-start issues in MF-based methods.

\subsection{Beneficial Applications}

Using our system, we can predict student grades in the next enrollment term much
more accurately than a random guessing strategy would. This information can be
used to aid students, educators, and advisors in a variety of ways. \textbf{For
students}, we can incorporate this information into a degree planning system.
Such systems already have ways of determining which courses meet which degree
requirements \cite{parameswaran_recommendation_2011}. Given several sets of
possible course selections for a semester, our system can be used to maximize
the expected grade. We could also provide students with a personalized
difficulty rating for each course based on their expected grade. This would help
students prioritize studies for each course and prepare properly for
particularly challenging courses.

\textbf{For educators}, knowledge of which students have the lowest expected
grades could provide opportunities to increase detection of at-risk students.
While students can seek help and have access to many resources, research shows
that at-risk students perform better if instructors are proactive in identifying
and reaching out to them \cite{grayson1998identifying}. A great deal of research
has gone into identifying characteristics of effective interventions based on
information provided by learning analytics. The Open Source Analytics Initiative
(OSAI) conducted a survey and analysis of this body of research, identifying
several key characteristics \cite{jayaprakash_early_2014}. Interventions that
increase student communication with instructors, connect students to available
university services, and provide students with an \textit{accurate assessment of
how they are currently performing} can be effective at increasing both retention
and academic performance. Our grade prediction system can be used as a component
of an early-warning system that can equip instructors to provide interventions
with these crucial characteristics.

The information provided by our system would also be invaluable \textbf{for
advisors}. Anticipating student performance and recommending the best plan of
action is a critical component of advisor responsibilities. Any additional
information that helps them personalize their advice to each student could
potentially help thousands of students. For instance, imagine a particular
course is known to be a challenging bottleneck course and another course is
known to be particularly easy in general. Without additional information, an
advisor might always recommend students take these two courses in the same
semester. However, some students might have trouble with exactly the type of
material taught or exactly the type of teaching style used in the ``easy''
course. Our system could provide advisors with this type of personalized
information.

\section{Conclusions and Future Work}

Motivated by the need for institutions to retain students, ensure timely
graduation, and ensure students are well-trained and workforce-ready in their
field of study, we have developed a system for next-term student grade
prediction. After experimenting with a wide variety of regression and
factorization models, we determined that a hybrid of the Random Forest model and
the MF-based Factorization Machine is best-suited to this task. Key to the
success of this hybrid model is the application of MADImp to FM feature
selection. Using this hybrid, we can predict grades for both new and returning
students and for both new and existing courses. While this system significantly
outperforms random guess predictions, and noticeably outperforms the other
regression models tested, it still has limitations we seek to address in future
works.

Of the methods surveyed, tensor factorization, multi-relational matrix
factorization, and custom tree-based methods seem most promising. The first two
methods are inspired by similar work in the ITS community. Both of these methods
hold promise for effectively leveraging the valuable instructor features
identified in this study. The idea of using more customized tree-based methods
is motivated by the desire for interpretable decision-making rules. Easily
understandable decision rules can speed student, educator, and advisor adoption
of new practices motivated by our findings. Additional feature engineering and
custom methods of constructing and pruning decision trees seem promising areas
for further work.

As in other MF applications, we observed the cold-start problem is a limiting
factor and complementary methods were required to overcome it. We overcame this
problem by combining FMs with RFs to compose a hybrid model. We have also
discussed using attribute-to-latent-feature mappings and identifying means of
overcoming covariate shift while learning 2-way interactions as two possible
augmentations to help FMs overcome the cold-start problem. Further research in
this area could greatly benefit many fields where MF-based methods are applicable.

Once we have improved predictive performance for both failing and passing
grades, we plan to deploy this system for live use as a component of a degree
planner and an early-warning system for students, instructors, and advisors.
Once live, we would perform A/B testing to better understand its performance,
the effect it has on the decision-making of its various users, and its ability
to improve student retention and learning success outcomes.

\section*{Acknowledgment}
\begin{footnotesize}
This research was funded by NSF IIS grant 1447489.
The data for this study was made available through a collaborative effort
spearheaded by Office of Institutional Research and Reporting and we would like
to acknowledge Kris Smith, Kathryn Zora, Angela Detlev, Eve Dauer, Kathy
Zimmerman, Joe Balducci, and Judy Lou at GMU.
\end{footnotesize}


\bibliographystyle{acmtrans}
\bibliography{refs}

\begin{thebibliography}{}

\bibitem[\protect\citeauthoryear{??}{gmu}{2014}]{gmu_facts_figures_2014}
 2014.
\newblock Gmu facts 2014-2015 facts and figures.

\bibitem[\protect\citeauthoryear{Adamic, Lukose, Puniyani, and Huberman}{Adamic
  et~al\mbox{.}}{2001}]{adamic_search_2001}
{\sc Adamic, L.~A.}, {\sc Lukose, R.~M.}, {\sc Puniyani, A.~R.}, {\sc and} {\sc
  Huberman, B.~A.} 2001.
\newblock Search in power-law networks.
\newblock {\em Physical Review E\/}~{\em 64,\/}~4.

\bibitem[\protect\citeauthoryear{Adomavicius and Tuzhilin}{Adomavicius and
  Tuzhilin}{2005}]{adomavicius_toward_2005}
{\sc Adomavicius, G.} {\sc and} {\sc Tuzhilin, A.} 2005.
\newblock Toward the next generation of recommender systems: A survey of the
  state-of-the-art and possible extensions.
\newblock {\em {IEEE} Trans. on Knowl. and Data Eng.\/}~{\em 17,\/}~6,
  734--749.

\bibitem[\protect\citeauthoryear{Alsumait, Wang, Domeniconi, and
  Barbara}{Alsumait et~al\mbox{.}}{2010}]{alsumait_embedding_2010}
{\sc Alsumait, L.}, {\sc Wang, P.}, {\sc Domeniconi, C.}, {\sc and} {\sc
  Barbara, D.} 2010.
\newblock Embedding semantics in {LDA} topic models.
\newblock In {\em Text Mining}, {M.~W. Berry} {and} {J.~Kogan}, Eds. John Wiley
  \& Sons, Ltd, 183--204.

\bibitem[\protect\citeauthoryear{{Arkadiusz Paterek}}{{Arkadiusz
  Paterek}}{2007}]{arkadiusz_paterek_improving_2007}
{\sc {Arkadiusz Paterek}}. 2007.
\newblock Improving regularized singular value decomposition for collaborative
  filtering.
\newblock {\em KDD\/}.

\bibitem[\protect\citeauthoryear{Aud and Wilkinson-Flicker}{Aud and
  Wilkinson-Flicker}{2013}]{NCES2013}
{\sc Aud, S.} {\sc and} {\sc Wilkinson-Flicker, S.} 2013.
\newblock {\em The Condition of Education 2013}.
\newblock Number NCES 2013-037. U.S. Department of Education, National Center
  for Education Statistics.

\bibitem[\protect\citeauthoryear{Baker, Corbett, and Aleven}{Baker
  et~al\mbox{.}}{2008}]{baker2008improving}
{\sc Baker, R.~S.}, {\sc Corbett, A.~T.}, {\sc and} {\sc Aleven, V.} 2008.
\newblock Improving contextual models of guessing and slipping with a truncated
  training set.
\newblock {\em Human-Computer Interaction Institute\/}, 17.

\bibitem[\protect\citeauthoryear{Barnes and Stamper}{Barnes and
  Stamper}{2008}]{barnes_toward_2008}
{\sc Barnes, T.} {\sc and} {\sc Stamper, J.} 2008.
\newblock Toward automatic hint generation for logic proof tutoring using
  historical student data.
\newblock In {\em Proc. of the 9th International Conference on Intelligent
  Tutoring Systems} (2008). 373--382.

\bibitem[\protect\citeauthoryear{Bayer}{Bayer}{2015}]{bayer_fastfm_2015}
{\sc Bayer, I.} 2015.
\newblock {fastFM}: A library for factorization machines.
\newblock {\em {arXiv}:1505.00641v2\/}.

\bibitem[\protect\citeauthoryear{Bell, Koren, and Volinsky}{Bell
  et~al\mbox{.}}{2007}]{bell_bellkor_2007}
{\sc Bell, R.~M.}, {\sc Koren, Y.}, {\sc and} {\sc Volinsky, C.} 2007.
\newblock The {BellKor} solution to the netflix prize.
\newblock {\em {AT}\&T Labs – Research\/}.

\bibitem[\protect\citeauthoryear{Breiman}{Breiman}{2001}]{breiman_random_2001}
{\sc Breiman, L.} 2001.
\newblock Random forests.
\newblock {\em Machine Learning\/}~{\em 45,\/}~1, 5--32.

\bibitem[\protect\citeauthoryear{Corbett and Anderson}{Corbett and
  Anderson}{1994}]{corbett_knowledge_1994}
{\sc Corbett, A.~T.} {\sc and} {\sc Anderson, J.~R.} 1994.
\newblock Knowledge tracing: Modeling the acquisition of procedural knowledge.
\newblock {\em User Modeling and User-Adapted Interaction\/}~{\em 4,\/}~4,
  253--278.

\bibitem[\protect\citeauthoryear{Council}{Council}{2001}]{NRC2001}
{\sc Council, N.~R.} 2001.
\newblock {\em Building a Workforce for the Information Economy}.
\newblock National Academies Press.

\bibitem[\protect\citeauthoryear{Desmarais and Baker}{Desmarais and
  Baker}{2011}]{desmarais_review_2011}
{\sc Desmarais, M.~C.} {\sc and} {\sc Baker, R. S. J.~d.} 2011.
\newblock A review of recent advances in learner and skill modeling in
  intelligent learning environments.
\newblock {\em User Modeling and User-Adapted Interaction\/}~{\em 22,\/}~1,
  9--38.

\bibitem[\protect\citeauthoryear{Elbadrawy, Studham, and {George
  Karypis}}{Elbadrawy et~al\mbox{.}}{2015}]{elbadrawy_personalized_2015}
{\sc Elbadrawy, A.}, {\sc Studham, R.~S.}, {\sc and} {\sc {George Karypis}}.
  2015.
\newblock Personalized multi-regression models for predicting students'
  performance in course activities.
\newblock {\em {LAK, '15}\/}.

\bibitem[\protect\citeauthoryear{Gantner, Drumond, Freudenthaler, Rendle, and
  Schmidt-Thieme}{Gantner et~al\mbox{.}}{2010}]{gantner_learning_2010}
{\sc Gantner, Z.}, {\sc Drumond, L.}, {\sc Freudenthaler, C.}, {\sc Rendle,
  S.}, {\sc and} {\sc Schmidt-Thieme, L.} 2010.
\newblock Learning attribute-to-feature mappings for cold-start
  recommendations.
\newblock In {\em Proc. of the 2010 {IEEE} International Conference on Data
  Mining} (2010). {ICDM} '10. {IEEE} Computer Society, 176--185.

\bibitem[\protect\citeauthoryear{Grayson, Miller, and Clarke}{Grayson
  et~al\mbox{.}}{1998}]{grayson1998identifying}
{\sc Grayson, A.}, {\sc Miller, H.}, {\sc and} {\sc Clarke, D.~D.} 1998.
\newblock Identifying barriers to help-seeking: a qualitative analysis of
  students' preparedness to seek help from tutors.
\newblock {\em British Journal of Guidance and Counselling\/}~{\em 26,\/}~2,
  237--253.

\bibitem[\protect\citeauthoryear{Huete, Fernández-Luna, de~Campos, and
  Rueda-Morales}{Huete et~al\mbox{.}}{2012}]{huete_using_2012}
{\sc Huete, J.~F.}, {\sc Fernández-Luna, J.~M.}, {\sc de~Campos, L.~M.}, {\sc
  and} {\sc Rueda-Morales, M.~A.} 2012.
\newblock Using past-prediction accuracy in recommender systems.
\newblock {\em Information Sciences\/}~{\em 199}, 78--92.

\bibitem[\protect\citeauthoryear{Jayaprakash, Moody, Lauría, Regan, and
  Baron}{Jayaprakash et~al\mbox{.}}{2014}]{jayaprakash_early_2014}
{\sc Jayaprakash, S.~M.}, {\sc Moody, E.~W.}, {\sc Lauría, E.~J.}, {\sc Regan,
  J.~R.}, {\sc and} {\sc Baron, J.~D.} 2014.
\newblock Early alert of academically at-risk students: An open source
  analytics initiative.
\newblock ~{\em 1,\/}~1, 6--47.

\bibitem[\protect\citeauthoryear{Kanagal, Ahmed, Pandey, Josifovski, Yuan, and
  Garcia-Pueyo}{Kanagal et~al\mbox{.}}{2012}]{kanagal_supercharging_2012}
{\sc Kanagal, B.}, {\sc Ahmed, A.}, {\sc Pandey, S.}, {\sc Josifovski, V.},
  {\sc Yuan, J.}, {\sc and} {\sc Garcia-Pueyo, L.} 2012.
\newblock Supercharging recommender systems using taxonomies for learning user
  purchase behavior.
\newblock {\em Proc. {VLDB} Endow.\/}~{\em 5,\/}~10, 956--967.

\bibitem[\protect\citeauthoryear{{Kevin Carey}}{{Kevin
  Carey}}{2005}]{kevin_carey_choosing_2005}
{\sc {Kevin Carey}}. 2005.
\newblock Choosing to improve: Voices from colleges and universities with
  better graduation rates.

\bibitem[\protect\citeauthoryear{Koren and Bell}{Koren and
  Bell}{2011}]{koren_advances_2011}
{\sc Koren, Y.} {\sc and} {\sc Bell, R.} 2011.
\newblock Advances in collaborative filtering.
\newblock In {\em Recommender Systems Handbook}, {F.~Ricci}, {L.~Rokach},
  {B.~Shapira}, {and} {P.~B. Kantor}, Eds. Springer, 145--186.

\bibitem[\protect\citeauthoryear{Linden, Smith, and York}{Linden
  et~al\mbox{.}}{2003}]{linden_amazon.com_2003}
{\sc Linden, G.}, {\sc Smith, B.}, {\sc and} {\sc York, J.} 2003.
\newblock Amazon.com recommendations: item-to-item collaborative filtering.
\newblock {\em {IEEE} Internet Computing\/}~{\em 7,\/}~1, 76--80.

\bibitem[\protect\citeauthoryear{Nedungadi and Smruthy}{Nedungadi and
  Smruthy}{2016}]{nedungadi_personalized_2016}
{\sc Nedungadi, P.} {\sc and} {\sc Smruthy, T.~K.} 2016.
\newblock Personalized multi-relational matrix factorization model for
  predicting student performance.
\newblock In {\em Intelligent Systems Technologies and Applications},
  {S.~Berretti}, {S.~M. Thampi}, {and} {P.~R. Srivastava}, Eds. Vol. 384.
  Springer International Publishing, 163--172.

\bibitem[\protect\citeauthoryear{Parameswaran, Venetis, and
  Garcia-Molina}{Parameswaran
  et~al\mbox{.}}{2011}]{parameswaran_recommendation_2011}
{\sc Parameswaran, A.}, {\sc Venetis, P.}, {\sc and} {\sc Garcia-Molina, H.}
  2011.
\newblock Recommendation systems with complex constraints: A course
  recommendation perspective.
\newblock {\em {ACM} Transactions on Information Systems ({TOIS})\/}~{\em
  29,\/}~4, 20.

\bibitem[\protect\citeauthoryear{Pardos, Wang, and Trivedi}{Pardos
  et~al\mbox{.}}{2012}]{pardos_real_2012}
{\sc Pardos, Z.~A.}, {\sc Wang, Q.~Y.}, {\sc and} {\sc Trivedi, S.} 2012.
\newblock The real world significance of performance prediction.
\newblock {\em International Educational Data Mining Society\/}.

\bibitem[\protect\citeauthoryear{Pedregosa, Varoquaux, Gramfort, Michel,
  Thirion, Grisel, Blondel, Prettenhofer, Weiss, Dubourg, Vanderplas, Passos,
  Cournapeau, Brucher, Perrot, and Duchesnay}{Pedregosa
  et~al\mbox{.}}{2011}]{scikit-learn}
{\sc Pedregosa, F.}, {\sc Varoquaux, G.}, {\sc Gramfort, A. e.~a.}, {\sc
  Michel, V.}, {\sc Thirion, B.}, {\sc Grisel, O.}, {\sc Blondel, M.}, {\sc
  Prettenhofer, P.}, {\sc Weiss, R.}, {\sc Dubourg, V.}, {\sc Vanderplas, J.},
  {\sc Passos, A.}, {\sc Cournapeau, D.}, {\sc Brucher, M.}, {\sc Perrot, M.},
  {\sc and} {\sc Duchesnay, E.} 2011.
\newblock Scikit-learn: Machine learning in {P}ython.
\newblock {\em JMLR\/}~{\em 12}, 2825--2830.

\bibitem[\protect\citeauthoryear{Peña-Ayala}{Peña-Ayala}{2014}]{pena-ayala_educational_2014}
{\sc Peña-Ayala, A.} 2014.
\newblock Educational data mining: A survey and a data mining-based analysis of
  recent works.
\newblock {\em Expert Systems with Applications\/}~{\em 41,\/}~4, 1432--1462.

\bibitem[\protect\citeauthoryear{Rendle}{Rendle}{2012}]{rendle_factorization_2012}
{\sc Rendle, S.} 2012.
\newblock Factorization machines with {libFM}.
\newblock {\em {ACM} TIST\/}~{\em 3,\/}~3.

\bibitem[\protect\citeauthoryear{Rhodes, Ung, Zundel, Herold, and
  Stahovich}{Rhodes et~al\mbox{.}}{2013}]{rhodes_using_2013}
{\sc Rhodes, N.}, {\sc Ung, M.}, {\sc Zundel, A.}, {\sc Herold, J.}, {\sc and}
  {\sc Stahovich, T.} 2013.
\newblock Using a lexical analysis of student's self-explanation to predict
  course performance.
\newblock In {\em Educational Data Mining 2013} (2013).

\bibitem[\protect\citeauthoryear{Romero, López, Luna, and Ventura}{Romero
  et~al\mbox{.}}{2013}]{romero_predicting_2013}
{\sc Romero, C.}, {\sc López, M.-I.}, {\sc Luna, J.-M.}, {\sc and} {\sc
  Ventura, S.} 2013.
\newblock Predicting students' final performance from participation in on-line
  discussion forums.
\newblock {\em Computers \& Education\/}~{\em 68}, 458--472.

\bibitem[\protect\citeauthoryear{Romero, Ventura, and García}{Romero
  et~al\mbox{.}}{2008}]{romero_data_2008}
{\sc Romero, C.}, {\sc Ventura, S.}, {\sc and} {\sc García, E.} 2008.
\newblock Data mining in course management systems: Moodle case study and
  tutorial.
\newblock {\em Computers \& Education\/}~{\em 51,\/}~1, 368--384.

\bibitem[\protect\citeauthoryear{Salakhutdinov, Mnih, and Hinton}{Salakhutdinov
  et~al\mbox{.}}{2007}]{salakhutdinov_restricted_2007}
{\sc Salakhutdinov, R.}, {\sc Mnih, A.}, {\sc and} {\sc Hinton, G.} 2007.
\newblock Restricted boltzmann machines for collaborative filtering.
\newblock In {\em Proc. of the 24th ICML}. ACM, 791--798.

\bibitem[\protect\citeauthoryear{Shan and Banerjee}{Shan and
  Banerjee}{2010}]{shan_generalized_2010}
{\sc Shan, H.} {\sc and} {\sc Banerjee, A.} 2010.
\newblock Generalized probabilistic matrix factorizations for collaborative
  filtering.
\newblock In {\em ICDM '10} (2010-12). 1025--1030.

\bibitem[\protect\citeauthoryear{Shani, Heckerman, and Brafman}{Shani
  et~al\mbox{.}}{2005}]{shani_mdp-based_2005}
{\sc Shani, G.}, {\sc Heckerman, D.}, {\sc and} {\sc Brafman, R.~I.} 2005.
\newblock An {MDP}-based recommender system.
\newblock {\em JMLR\/}~{\em 6}, 1265--1295.

\bibitem[\protect\citeauthoryear{Shimodaira}{Shimodaira}{2000}]{shimodaira_improving_2000}
{\sc Shimodaira, H.} 2000.
\newblock Improving predictive inference under covariate shift by weighting the
  log-likelihood function.
\newblock {\em Journal of Statistical Planning and Inference\/}~{\em 90,\/}~2,
  227--244.

\bibitem[\protect\citeauthoryear{Sorour, Mine, Goda, and Hirokawa}{Sorour
  et~al\mbox{.}}{2015}]{sorour_predictive_2015}
{\sc Sorour, S.~E.}, {\sc Mine, T.}, {\sc Goda, K.}, {\sc and} {\sc Hirokawa,
  S.} 2015.
\newblock A predictive model to evaluate student performance.
\newblock {\em Journal of Information Processing\/}~{\em 23,\/}~2, 192--201.

\bibitem[\protect\citeauthoryear{Sweeney, {Lester, Jaime}, and {Rangwala,
  Huzefa}}{Sweeney et~al\mbox{.}}{2015}]{sweeney_next-term_2015}
{\sc Sweeney, M.}, {\sc {Lester, Jaime}}, {\sc and} {\sc {Rangwala, Huzefa}}.
  2015.
\newblock Next-term student grade prediction.
\newblock {\em {IEEE} Big Data 2015\/}.

\bibitem[\protect\citeauthoryear{Thai-Nghe, Drumond, Horváth, Nanopoulos, and
  Schmidt-Thieme}{Thai-Nghe et~al\mbox{.}}{2011}]{thai-nghe_matrix_2011}
{\sc Thai-Nghe, N.}, {\sc Drumond, L.}, {\sc Horváth, T.}, {\sc Nanopoulos,
  A.}, {\sc and} {\sc Schmidt-Thieme, L.} 2011.
\newblock Matrix and tensor factorization for predicting student performance.
\newblock In {\em {CSEDU} (1)} (2011). 69--78.

\bibitem[\protect\citeauthoryear{Thai-Nghe, Drumond, Horváth, and
  Schmidt-Thieme}{Thai-Nghe et~al\mbox{.}}{2012}]{thai-nghe_using_2012}
{\sc Thai-Nghe, N.}, {\sc Drumond, L.}, {\sc Horváth, T.}, {\sc and} {\sc
  Schmidt-Thieme, L.} 2012.
\newblock Using factorization machines for student modeling.
\newblock In {\em {UMAP} Workshops} (2012).

\bibitem[\protect\citeauthoryear{Thai-Nghe, Drumond, Horváth, Schmidt-Thieme,
  and {others}}{Thai-Nghe
  et~al\mbox{.}}{2011}]{thai-nghe_multi-relational_2011}
{\sc Thai-Nghe, N.}, {\sc Drumond, L.}, {\sc Horváth, T.}, {\sc
  Schmidt-Thieme, L.}, {\sc and} {\sc {others}}. 2011.
\newblock Multi-relational factorization models for predicting student
  performance.
\newblock In {\em {KDD} Workshop on Knowledge Discovery in Educational Data
  ({KDDinED})} (2011).

\bibitem[\protect\citeauthoryear{Tinto}{Tinto}{2006}]{tinto_research_2006}
{\sc Tinto, V.} 2006.
\newblock Research and practice of student retention: what next?
\newblock {\em Journal of College Student Retention: Research, Theory \&
  Practice\/}~{\em 8,\/}~1, 1--19.

\bibitem[\protect\citeauthoryear{Xu and Mostow}{Xu and
  Mostow}{2012}]{xu_comparison_2012}
{\sc Xu, Y.} {\sc and} {\sc Mostow, J.} 2012.
\newblock {\em Comparison of Methods to Trace Multiple Subskills: Is {LR}-{DBN}
  Best?}
\newblock International Educational Data Mining Society.

\bibitem[\protect\citeauthoryear{Xu and Mostow}{Xu and
  Mostow}{2015}]{xu_unified_2015}
{\sc Xu, Y.} {\sc and} {\sc Mostow, J.} 2015.
\newblock A unified 5-dimensional framework for student models.
\newblock {\em International Educational Data Mining Society\/}.

\end{thebibliography}

\section*{Appendix I: Feature Descriptions} \label{feature-descriptions}
\begin{small}

This section provides detailed descriptions of each feature.

\subsection{Student Features}

\begin{itemize}
    \itemsep0em
    \item  \textit{sid}: Unique identifier of the student. When used in training
        data, the student ID is one-hot encoded to learn student bias terms.
    \item  \textit{grdpts}: [0, 4] grade the student has obtained for a particular course.
    \item  \textit{major}: Declared major during current term.
    \item  \textit{race}: Self-reported race of student; may be unspecified.
    \item  \textit{sex}: Self-reported gender; may be unspecified.
    \item  \textit{age}: Age determined from birth date in admissions records.
    \item  \textit{zip}: Zip code, or postal code for students from outside the US.
    \item  \textit{sat}: 1600-scale SAT score, if available.
    \item  \textit{hs}: High school CEEB code.
    \item  \textit{hsgpa}: High school GPA. For transfer students, this feature
        contains the GPA from the institution the student is transferring from.
    \item  \textit{lterm\_gpa}: Term GPA from the previous term.
    \item  \textit{lterm\_cum\_gpa}: Cumulative GPA as of the previous term.
    \item  \textit{term\_chrs}: Number of credit hours the student is enrolled in during
        the current term.
    \item  \textit{total\_chrs}: Number of credit hours student has taken (not passed) up
        to the previous term.
    \item  \textit{alevel}: Academic level of the student. Obtained by binning
        total\_chrs: [0,30)=0, [30,60)=1, [60,90)=2, [90,120]=3, (120+)=4.
    \item  \textit{sterm}: Chronological numbering of terms relative to this
        student. The student's first term is 0, his second is 1, and so on.
\end{itemize}

\subsection{Course Features}

\begin{itemize}
    \itemsep0em
    \item  \textit{cid}: Unique identifier of the course. When used in training
        data, the student ID is one-hot encoded to learn course bias terms.
    \item  \textit{cdisc}: Course discipline.
    \item  \textit{chrs}: Number of credit hours this course is worth.
    \item  \textit{clevel}: Course level [1, 7].
    \item  \textit{termnum}: Number of the term this course was offered in.
        These are relative to the dataset only. Term 0 is the first term for
        which we have data, and they are numbered chronologically onwards from
        there.
    \item  \textit{num\_enrolled}: Number of students enrolled in this course
        for the current term, across all sections.
    \item  \textit{total\_enrolled}: Total number of students enrolled in this
        course since its first offering, including the current term.
    \item  \textit{lterm\_cgpa}: The aggregate [0, 4] GPA of all students who
        took this course during the previous term.
    \item  \textit{lterm\_cum\_cgpa}: The aggregate [0, 4] GPA of all students
        who have ever taken this class up to the previous term.
\end{itemize}

\subsection{Instructor Features}

\begin{itemize}
    \itemsep0em
    \item  \textit{iid}: Unique identifier of the instructor. When used in
        training data, the student ID is one-hot encoded to learn instructor
        bias terms.
    \item  \textit{iclass}: Classification (Adjunct, Full time, Part time, GRA,
        GTA).
    \item  \textit{irank}: Rank (Instructor, Assistant Professor, Associate
        Professor, Eminent Scholar, University Professor).
    \item  \textit{itenure}: Tenure status (Term, Tenure-track, Tenured).
\end{itemize}

\end{small}

\section*{Appendix II: Model Parameter Settings} \label{parameter-settings}
\begin{small}
\begin{itemize}
    \item  \textit{Factorization Machine}: k = rank = 8, i = number of
        iterations = 200, s = initial standard deviation = 0.2
    \item  \textit{Personalized Multi-Linear Regression}: k = number of models =
        4, lw = regularization on $P$ and $W$ = 0.01, lb = $\lambda_B$ = 0.5, lr
        = SGD learning rate = 0.001
    \item  \textit{Random Forest}: n = number of trees = 100, m = max depth = 10
    \item  \textit{Boosted Decision Trees}: n = number of trees = 100, m = max
        depth = 11
    \item  \textit{Stochastic Gradient Descent (SGD) Regression}:
        lr = learning rate = 0.001, r = regularization term = 0.001, i =
        number of iterations = 15
    \item  \textit{Ordinary Least Squares (OLS) Regression}: no parameters
    \item  \textit{k-Nearest Neighbors (kNN)}: k = number of neighbors = 20
    \item  \textit{Decision Tree}: m = max depth = 4
\end{itemize}
\end{small}

\section*{Appendix III: MADImp Example} \label{madimp-example}
\begin{small}

Let us consider a particular example to better understand the idea of Mean
Absolute Deviation Importance (MADImp). We proceed with the FM model but note
again that MADImp can be used with any generalized linear model (GLM).

Assume we have three features: a user ID, an item ID, and a season (Spring,
Summer, Fall, Winter). We one-hot encode all features, so for a particular
record we have three features with a value of 1 and all the rest have values of
0. Now the goal is to predict the response a particular user gives to a
particular item during a particular season. Let us now fixate upon a single
user. This user generally tends to give all items a slightly positive response.
The responses are more positive in the Spring and Fall, and less positive in the
Summer and Winter. Next we fixate upon a single item. This item usually
generates very positive responses, with the positivity spiking in the Summer.
The other seasons have no effect on users' responses to the item. The user has
no particular preferences for or against the item. We also observe the overall
pattern of responses is slightly positive and the range of ratings is from 0 to
4. Finally, we do not observe any general seasonality effects. We only find
seasonal preferences of and for particular users and items.

Given this setting, we can now assign hypothetical values to our parameters. We
have 10 users and 20 items, and there are 4 seasons. So after one-hot encoding,
we have a total of 34 features. Let our user be feature 1, our item be feature
11, and the seasons are then the following feature numbers: Spring = 31, Summer
= 32, Fall = 33, Winter = 34. As is typical with regression models, we set
feature 0 to 1 for all records to learn the global intercept term.

\begin{itemize}
    \item  $w_0    = 0.5$, a positive global intercept term.
    \item  $w_1    = 0.5$, a positive reponse trend for this user.
    \item  $w_{11} = 2.0$, a very positive response trend for this item.
    \item  $Z_{1,31} = Z_{1,33} =  0.2$, this user has slightly more positive
        responses in Spring and Fall.
    \item  $Z_{1,32} = Z_{1,34} = -0.2$, this user has slightly less positive
        responses in Summer and Winter.
    \item  $Z_{11,32} = 0.2$, this item has slightly more positive responses in
        the Summer.
\end{itemize}

With these parameter values, a prediction for this user-item combo in the Summer
would be calculated as:

\begin{align}
    \hat{y}(X_d) \begin{aligned}[t]
         &= w_0 + w_1 + w_{11} + Z_{1,32} + Z_{11,32}  \\
         &= 0.5 + 0.5 + 2.0 - 0.2 + 0.2  \\
         &= 3.0
    \end{aligned}
    \nonumber
\end{align}

Recall that we have three features set to 1 for each dyad feature vector $X_d$.
So we calculate the importance of these three features using
(\ref{eq:importance-single-dyad}).
\begin{align*}
    T_d
        &= |w_1| + |w_{11}| + |Z_{1,32}| + |Z_{11,32}|  \\
        &= |0.5| + |2.0| + |-0.2| + |0.2| = 2.9  \\[3mm]
    I(X_{d,1})
        &= \frac{|w_1| + |Z_{1,32}| / 2}{T_d}  \\
        &= \frac{|0.5| + |-0.2| / 2}{2.9}
        = \frac{0.6}{2.9} \approx 0.2069  \\[3mm]
    I(X_{d,11})
        &= \frac{|w_{11}| + |Z_{11,32}| / 2}{T_d}  \\
        &= \frac{|2.0| + |0.2| / 2}{2.9}
        = \frac{2.1}{2.9} \approx 0.7241  \\[3mm]
    I(X_{d,32})
        &= \frac{|Z_{1,32}| / 2 + |Z_{11,32}| / 2}{T_d}  \\
        &= \frac{|-0.2| / 2 + |0.2| / 2}{2.9}
         = \frac{0.2}{2.9} \approx 0.0690  \\
\end{align*}
\end{small}

\end{document}